\documentstyle[epsf]{elsart}

\def\lsim{\mathrel{\vcenter{\hbox{$<$}\nointerlineskip\hbox{$\sim$}}}}
\def\gsim{\mathrel{\vcenter{\hbox{$>$}\nointerlineskip\hbox{$\sim$}}}}

\begin{document}

\begin{frontmatter}

\hfill HIP-1999-16/TH

\title{Phenomenology of light Higgs bosons in  supersymmetric 
left-right models}

\author{K. Huitu$^a$, P. N. Pandita$^{a,b,c}$ and
K. Puolam\"{a}ki$^a$}

\address{$^a$Helsinki Institute of Physics, P.O.Box 9, FIN-00014
University of Helsinki, Finland}

\address{$^b$ Deutsches Elektronen-Synchrotron DESY, 
Notkestrasse 85, D-22603 Hamburg, Germany}

\address{$^c$Department of Physics, North Eastern Hill
University, Shillong 793022, India\thanksref{pran}}
\thanks[pran]{Permanent address}

\begin{abstract}

We carry out a detailed analysis of the light Higgs bosons in 
supersymmetric left-right models (SLRM). 
This includes models with minimal particle content and those with additional
Higgs superfields.
We also consider models with non-renormalizable 
higher-dimensional terms.
We obtain an upper bound on the mass of the lightest $CP$-even neutral 
Higgs boson in these models. 
The upper bound  depends only on the gauge couplings, and the vacuum 
expectation values of those neutral Higgs fields which control the
spontaneous breakdown of the $SU(2)_L \times U(1)_Y$ gauge symmetry. 
We calculate the one-loop radiative corrections to this upper bound,
and evaluate it numerically in the minimal version of 
the supersymmetric left-right model.
We consider the couplings of this  lightest $CP$-even Higgs boson 
to the  fermions, and 
show that in a phenomenologically viable model the branching ratios
are similar to the corresponding branching ratios in the minimal
supersymmetric standard model (MSSM).
We then study the most promising particle 
for distinguishing the SLRM from other models, namely the 
doubly charged Higgs boson.
We obtain the mass of this doubly charged Higgs boson
in different types of supersymmetric left-right models,
and discuss its phenomenology.

\end{abstract}


\end{frontmatter}

\section{Introduction}

One of the central problems of particle physics is to understand how
the electroweak scale associated with the mass of the W boson is
generated, and why it is so small as compared to the Planck scale
associated with the Newton's constant. In the Standard Model (SM) the
electroweak scale is generated through the vacuum expectation value
(VEV) of the neutral component of a  Higgs doublet
\cite{Gunion:1989we}. Apart from the fact that this VEV is an arbitrary
parameter in the SM, the mass parameter of the Higgs field suffers
from quadratic divergences, making the weak scale unstable under
radiative corrections. Supersymmetry is at present the only known
framework in which the weak scale is stable under radiative
corrections \cite{Haber:1985rc}, although it 
does not explain 
how such a small scale arises in the first place. As such, 
considerable importance attaches to the study of supersymmetric
models, especially the Minimal Supersymmetric Standard Model (MSSM),
based on the gauge group $SU(2)_L \times U(1)_Y$, with two Higgs
doublet superfields. It is well known that, because of underlying
gauge invariance and supersymmetry (SUSY), the lightest Higgs boson in 
MSSM has a tree level upper bound of $M_Z$ (the mass of Z boson) on
its mass \cite{Inoue:1982ej}. Although radiative corrections 
\cite{Okada:1991vk}
to the tree level result can be appreciable, these depend only
logarithmically on the SUSY breaking scale, and are, therefore, under
control. This results in an upper bound of about $125-135$ GeV on the
one-loop radiatively corrected mass \cite{Rosiek:1995dy}
of the lightest Higgs boson in MSSM\footnote{The two-loop corrections
to the Higgs boson mass matrix in the MSSM are significant, and can reduce
the lightest 
Higgs boson mass by up to $\sim 20$ GeV as compared to its one-loop 
value \cite{twoloop}.}. 
Because of the presence of the additional trilinear
Yukawa couplings, such a tight constraint on the mass of the lightest
Higgs boson need not {\em a priori} hold in extensions of MSSM 
based on the gauge group $SU(2)_L \times U(1)_Y$ with an
extended Higgs sector. Nevertheless, it has been shown that the upper
bound on the lightest Higgs boson mass in these models depends only on
the weak scale, and dimensionless coupling constants (and only
logarithmically on the SUSY breaking scale), and is calculable if all
the couplings remain perturbative below some scale $\Lambda$
\cite{Ellis:1989er,Binetruy:1992mk,Pandita:1993hx,Elliott:1993ex,Ellwanger:1993hn,Kane:1993kq,EQ}. 
This upper
bound can vary between $150$ GeV to $200$ GeV depending on the Higgs
structure of the underlying supersymmetric model. Thus, nonobservation
of such a light Higgs boson below this upper bound will rule out an entire
class of supersymmetric models based on the gauge group $SU(2)_L
\times U(1)_Y$.

The existence of the upper bound on the lightest Higgs boson mass in MSSM
(with arbitrary Higgs sectors) has been investigated in a situation
where the underlying supersymmetric model respects baryon ($B$) and
lepton ($L$) number conservation. However, it is well known that gauge
invariance, supersymmetry and renormalizibility allow $B$ and $L$
violating terms in the superpotentioal of the MSSM
\cite{Weinberg:1982wj}. 
The strength of these lepton and baryon number violating terms is,
however, severely limited by phenomenological
\cite{Zwirner:1984is,Smirnov:1996ey},  
and cosmological
\cite{Campbell:1991fa} constraints. 
Indeed, unless the strength of
baryon-number violating term is less than $10^{-13}$,
 it will lead to contradiction with the present
lower limits on the lifetime of the proton. The usual strategy to
prevent the appearance of $B$ and $L$ violating couplings in MSSM is
to invoke a discrete $Z_2$ symmetry 
\cite{Farrar:1978xj} 
known
as matter parity, or R-parity. The matter parity of each superfield may
be defined as
\begin{equation}
(\rm{matter\; parity}) \equiv (-1)^{3 (B-L)} .
\label{eq:parity}
\end{equation}
The multiplicative conservation of matter parity forbids all the
renormalizable $B$ and $L$ violating terms in the superpotential of
MSSM. Equivalently, the R-parity of any component {\em field} is
defined by $R_p = (-1)^{3(B-L)+2S}$, where $S$ is the spin of the
field. Since $(-1)^{2S}$ is conserved in any Lorentz-invariant
interaction, matter parity conservation and R-parity conservation are
equivalent. Conservation of R-parity  then immediately implies
that superpartners can be produced only in pairs, and that the
lightest supersymmetric particle (LSP) is absolutely stable.

Although the Minimal Supersymmetric Standard Model with R-parity
conservation can provide a description of nature which is consistent
with all known observations, the assumption of $R_p$ conservation
appears to be {\em ad hoc}, since it is not required for the internal
consistency of MSSM. Furthermore, all global symmetries, discrete or
continuous, could be violated by the Planck scale physics effects
\cite{Giddings:1988cx}. The problem becomes acute for low energy
supersymmetric models, because $B$ and $L$ are no longer automatic
symmetries of the Lagrangian, as they are in the Standard Model.
It is, therefore, more appealing to have a supersymmetric theory
where R-parity is related to a gauge symmetry, and its conservation is
automatic because of the invariance of the underlying theory
under this gauge symmetry. Fortunately, there is a compelling 
scenario which does
provide for exact R-parity conservation due to a deeper
principle. Indeed, $R_p$ conservation follows automatically in certain
theories with gauged $(B-L)$, as is suggested by the fact that matter
parity is simply a $Z_2$ subgroup of $(B-L)$. It has been noted by
several authors 
\cite{Mohapatra:1986su,Martin:1992mq} 
that if the gauge symmetry
of MSSM is extended to $SU(2)_L \times U(1)_{I_{3R}} \times
U(1)_{B-L}$, or $SU(2)_L \times SU(2)_R \times U(1)_{B-L}$, the theory
becomes automatically R-parity conserving. Such a left-right
supersymmetric theory (SLRM) solves the problems of explicit $B$ and
$L$ violation of MSSM, and has received much attention recently
\cite{Cvetic:1984su,Francis:1991pi,Kuchimanchi:1993jg,Huitu:1994gf,Huitu:1995zm,Huitu:1997iy,Aulakh:1997ba}.
Of course left-right symmetric theories 
are also interesting in their own right,
for among other appealing features, they offer a simple and natural
explanation for the  smallness of neutrino mass through the 
so called see-saw
mechanism \cite{Gell-Mann:1979}. 

Such a naturally R-parity conserving theory necessarily involves the
extension of the Standard Model gauge group, and since the extended
gauge symmetry has to be broken, it involves a ``new scale'', the
scale of left-right symmetry breaking, beyond the SUSY and $SU(2)_L
\times U(1)_Y$ breaking scales of MSSM. 
In \cite{Huitu:1997rr} we showed that in the SLRM with minimal particle
content  the upper  bound on  the mass of the 
lightest neutral Higgs boson depends
only on gauge couplings and those VEVs which break 
the $SU(2)_L\times U(1)_Y$
symmetry.  Here we will present a detailed analysis  
of the Higgs sector of the left-right 
supersymmetric models, and consider 
some of the distinguishing features of the lightest Higgs boson
in these models.
In the SLRMs there are typically also other light Higgs particles.
The light doubly charged Higgs boson, which we will consider in
detail as well,
should provide a clear signal in experiments.

The plan of the paper is as follows. In section \ref{sec:tree}, 
we review the various left-right supersymmetric models that 
we consider in this paper.
In section \ref{sec:higgs}, we obtain the tree level upper bound on
the mass of the lightest CP-even Higgs boson for the various SLRMs
considered in section \ref{sec:tree}.
We use  a general 
procedure to obtain this  (tree-level) upper bound 
on the mass of the lightest CP-even Higgs boson 
in models with extended gauge groups, such as SLRMs.  
We show that the upper bound so obtained 
in the renormalizable models is independent of the 
supersymmetry breaking scale, as well as the left-right 
breaking scale. In the case of models containing non-renormalizable 
terms, although the upper bound depends on the left-right 
breaking scale, the dependence is extremely weak, being 
suppressed by powers of Planck mass.

In section \ref{sec:rad} we calculate the radiative corrections to the
upper bound on the mass of the lightest Higgs boson, and show that the most
important radiative corrections arising from quark-squark loops are of
the same type as in the MSSM based on $SU(2)_L \times
U(1)_Y$. 
The radiatively corrected upper bound so obtained is 
numerically considerably larger than
the corresponding bound in the MSSM, but for most of the parameter
space is below 200 GeV.
In section \ref{sec:decoup} we consider the branching ratios of the
lightest neutral Higgs boson,
and find that its couplings are similar to the corresponding
Higgs couplings in the Standard Model in the decoupling limit.

In section \ref{sec:double} we discuss the mass of the lightest doubly
charged Higgs boson and the possibility of 
its detection at colliders.
In section \ref{sec:conclusion}, we present our conclusions. 
The full scalar potential of the minimal supersymmetric
left-right model is presented
in the Appendix A.

\section{The Higgs sector of the left-right supersymmetric models} 
\label{sec:tree}

In this section we briefly review the minimal supersymmetric left-right model, 
and then discuss models which have an extended Higgs sector, and 
finally discuss models with non-renormalizable interaction
terms in the superpotential.

The minimal SLRM is based on the gauge
group $SU(3)_C \times SU(2)_L \times SU(2)_R \times U(1)_{B-L}$. The
matter fields of this model consist of the three families of quark
and lepton chiral superfields with the following transformation
properties under the gauge group: 
\begin{eqnarray}
Q=\left( \begin{array}{c} U \\ D \end{array} \right) \sim
(3,2,1,\frac{1}{3}) & ,\; & Q^c = \left( \begin{array}{c} D^c \\ U^c
\end{array} \right) \sim (3^*,1,2,-\frac{1}{3}) , \nonumber \\ L=\left(
\begin{array}{c} \nu \\ E \end{array} \right) \sim 
(1,2,1,-1) & ,\; & L^c =\left( \begin{array}{c} E^c \\ \nu^c \end{array}
\right) \sim (1,1,2,1) ,
\label{eq:fields1}
\end{eqnarray}
where the numbers in the brackets denote the quantum numbers under
$SU(3)_C \times SU(2)_L \times SU(2)_R \times U(1)_{B-L}$.  The Higgs
sector consists of the bidoublet and triplet Higgs superfields:
\begin{eqnarray}
\Phi = \left( \begin{array}{cc} \Phi^0_1 & \Phi^+_1 \\ \Phi^-_2 &
\Phi^0_2 \end{array} \right) \sim (1,2,2,0) & ,\; & \chi = \left(
\begin{array}{cc} \chi^0_1 & \chi^+_1 \\ \chi^-_2 & \chi^0_2
\end{array} \right) \sim (1,2,2,0) , \nonumber \\ \Delta_R = \left(
\begin{array}{cc} \frac{1}{\sqrt{2}} \Delta_R^- & \Delta_R^0 \\
\Delta_R^{--} & -\frac{1}{\sqrt{2}} \Delta^-_R \end{array} \right)
\sim (1,1,3,-2) & ,\; & \delta_R = \left(
\begin{array}{cc} \frac{1}{\sqrt{2}} \delta_R^+ & \delta_R^{++} \\
\delta_R^0 & -\frac{1}{\sqrt{2}} \delta^+_R \end{array} \right)
\sim (1,1,3,2)  , \nonumber \\ \Delta_L = \left(
\begin{array}{cc} \frac{1}{\sqrt{2}} \Delta_L^- & \Delta_L^0 \\
\Delta_L^{--} & -\frac{1}{\sqrt{2}} \Delta^-_L \end{array} \right)
\sim (1,3,1,-2) & ,\; & \delta_L = \left(
\begin{array}{cc} \frac{1}{\sqrt{2}} \delta_L^+ & \delta_L^{++} \\
\delta_L^0 & -\frac{1}{\sqrt{2}} \delta^+_L \end{array} \right) \sim
(1,3,1,2) .
\label{eq:fields2}
\end{eqnarray}
There are two bidoublet superfields 
in order to implement the $SU(2)_L
\times U(1)_Y$ breaking, and to generate a nontrivial Kobayashi-Maskawa
matrix. Furthermore, two $SU(2)_R$ Higgs triplet superfields
$\Delta_R$ and $\delta_R$ with opposite $(B-L)$ are necessary to break
the left-right symmetry spontaneously, and to cancel triangle gauge
anomalies due to the fermionic superpartners. The gauge symmetry is
supplemented by a discrete left-right symmetry under which
the fields can be chosen to transform as
\begin{eqnarray}
Q \leftrightarrow Q^c , \; L \leftrightarrow L^c , \;
\Phi \leftrightarrow -\tau_2 \Phi^T \tau_2 , \; \chi \leftrightarrow
-\tau_2 \chi^T \tau_2 , \; \Delta_R \leftrightarrow \delta_L , \;
\delta_R \leftrightarrow \Delta_L .
\end{eqnarray}
Thus, the $SU(2)_L$ triplets $\Delta_L$ and $\delta_L$ are needed in
order to make the Lagrangian fully symmetric under $L \leftrightarrow
R$ transformation, although these are not needed phenomenologically
for symmetry breaking, or the see-saw mechanism.

The most general gauge invariant superpotential involving these
superfields can be written as (generation indices suppressed)
\begin{eqnarray}
W_{min}&=& h_{\Phi Q}Q^T i\tau_2 \Phi Q^c + h_{\chi Q}Q^T i\tau_2 \chi Q^c +
h_{\Phi L}L^T i\tau_2 \Phi L^c + h_{\chi L}L^T i\tau_2 \chi L^c
\nonumber\\ &&+h_{\delta_L} L^T i\tau_2 \delta_L L + h_{\Delta_R}
L^{cT} i\tau_2 \Delta_R L^c+ \mu_1 \rm{Tr} (i\tau_2\Phi^T i\tau_2
\chi) + \mu_1' \rm{Tr} (i\tau_2\Phi^T i\tau_2 \Phi) \nonumber\\ &&+
\mu_1'' \rm{Tr} (i\tau_2\chi^T i\tau_2 \chi) +\rm{Tr}
(\mu_{2L}\Delta_L \delta_L + \mu_{2R}\Delta_R\delta_R).
\label{eq:superpotential}
\end{eqnarray}
The scalar potential  can be calculated from
\begin{equation}
V=V_F+V_D+V_{soft},
\label{eq:scalarpotential}
\end{equation}
where $V_F$, $V_D$, and $V_{soft}$ represent the contribution
of $F$-terms, the $D$-terms, and the soft SUSY breaking terms,
respectively. The full scalar potential can be found in Appendix A.
The general form of the vacuum expectation values of the various
scalar fields which preserve the $U(1)_{\rm{em}}$ 
gauge invariance can be written as
\begin{eqnarray}
&&\langle \Phi \rangle = \left( \begin{array}{cc} \kappa_1 & 0 \\ 0 &
e^{i \phi_1}\kappa_1' \end{array} \right) ,\;  
\langle \chi \rangle =\left(\begin{array}{cc} e^{i \phi_2} \kappa_2' &
0 \\ 0 & \kappa_2\end{array} \right) , \;
\langle \Delta_R \rangle = \left(\begin{array}{cc} 0 & v_{\Delta_R} \\
0 & 0 \end{array} \right) ,\;  
\langle \delta_R \rangle = \left(\begin{array}{cc} 0 & 0 \\
v_{\delta_R} & 0 \end{array} \right) , \nonumber \\ 
&& \langle \Delta_L \rangle = \left(\begin{array}{cc} 0 & v_{\Delta_L} \\
0 & 0 \end{array} \right)  ,\;  
\langle \delta_L \rangle = \left(\begin{array}{cc} 0 & 0 \\
v_{\delta_L} & 0\end{array} \right) , \;
\langle L \rangle =\left(\begin{array}{c} \sigma_L \\ 0 \end{array}
\right)  ,\;  
\langle L^c\rangle = \left( \begin{array}{c} 0 \\ \sigma_R \end{array}
\right) .
\label{eq:vevs}
\end{eqnarray}
We note that the triplet vacuum expectation values $v_{\Delta_R}$ and
$v_{\delta_R}$ represent the scale of $SU(2)_R$ breaking and are,
according to the lower bounds \cite{Caso} on heavy
W- and Z-boson masses, in the range $v_{\Delta_R},v_{\delta_R} \gsim 1$
TeV. These represent a new scale, the right-handed breaking 
scale. We note that $\kappa_1'$
and $\kappa_2'$ contribute to the mixing of the charged gauge bosons
and to the flavour changing neutral currents, and are usually assumed
to vanish. Furthermore, since the electroweak $\rho$
parameter is
close to unity, $\rho = 0.9998 \pm 0.0008$ \cite{Caso},  
the triplet vacuum expectation values $v_{\Delta_L}$ and $v_{\delta_L}$
must be small.

The Yukawa coupling $h_{\chi L}$ is proportional to the neutrino Dirac
mass $m_D$. The light neutrino mass in the see-saw mechanism is given
by $\sim m_D^2 /m_M$, where $m_M = h_{\Delta_R} v_{\Delta_R}$ is the
Majorana mass. The magnitude of
$h_{\chi L}$ is not accurately  determined given the present
upper limit on the light neutrino masses. On the other hand the Yukawa
coupling $h_{\Phi L}$ is proportional to the electron mass and is,
thus, small.

In the minimal model described above, parity cannot be spontaneously broken
at the renormalizable level without spontaneous
breaking of $R$-parity. This may be cured  by adding more fields to
the theory. In \cite{Kuchimanchi:1993jg,Kuchimanchi:1995vk}
it was suggested that a parity-odd singlet,
coupled appropriately to triplet fields,  be introduced so as to ensure
proper symmetry breaking. This leads to a set of degenerate minima connected
by a flat direction, all of them breaking parity. When soft SUSY breaking 
terms are switched on, the degeneracy is lifted, but the global minimum
that results breaks $U(1)_{\rm{em}}$. Because of the flat direction 
connecting the minima, there is no hope that the fields remain in the 
phenomenologically
acceptable vacuum, which rolls down to global minimum after SUSY is softly 
broken. The only option that is left is to have a relatively 
low $SU(2)_R$ breaking scale,
with spontaneously broken $R$-parity ($\langle \nu^c \rangle 
\equiv \sigma_R$ is non-zero). We note that 
present experiments allow for a low $SU(2)_R$ breaking scale.

There is an alternative to the minimal left-right supersymmetric
model which
involves the addition of a couple of triplet fields, 
$\Omega_L(1,3,1,0)$ and $\Omega_R(1,1,3,0)$, instead of
a singlet Higgs superfield, to the minimal model \cite{Aulakh:1997ba}.
In these extended models the breaking of $SU(2)_R$ is achieved in two stages.
In the first stage
the gauge group $SU(2)_L \times SU(2)_R \times U(1)_{B-L}$ is
broken to an intermediate symmetry group
$SU(2)_L \times U(1)_R \times U(1)_{B-L}$, and at the
second stage  $U(1)_R \times U(1)_{B-L}$ is broken to $U(1)_Y$ at
a lower scale. In this theory there is only one
parity-breaking minimum, in contrast to the minimal model, which
respects the electromagnetic gauge invariance.
The superpotential for this class of models
obtains  additional terms involving the triplet fields
$\Omega_L$ and $\Omega_R$: 
\begin{eqnarray}
 W_{\Omega}&=& W_{min} + 
 \frac 12 \mu_{\Omega_L}   {\rm{Tr}}\,\Omega_L^2
	  +\frac 12 \mu_{\Omega_R}  {\rm Tr}\,\Omega_R^{2} +a_L {\rm Tr}\,\Delta_L \Omega_L \delta_L
       +a_R {\rm Tr}\,\Delta_R \Omega_R \delta_R \nonumber \\
& &  + {\rm Tr}\, \Omega_L  \left( \alpha_L \Phi i\tau_2 \chi^T
 i\tau_2+{\alpha_L}' \Phi i\tau_2 \Phi^T i\tau_2 +{\alpha_L}'' \chi i\tau_2
 \chi^T i\tau_2  \right)
 \nonumber \\ & & + {\rm Tr}\, \Omega_R \left( \alpha_R i \tau_2 \Phi^T i\tau_2
 \chi+{\alpha_R}' i\tau_2 \Phi^T i\tau_2 \Phi +{\alpha_R}'' i\tau_2\chi^T i\tau_2
 \chi  \right),
\label{superpot}
\end{eqnarray}
where $W_{min}$ is the superpotential (\ref{eq:superpotential}) of the  minimal
left-right model.
In these models the see-saw mechanism takes its canonical
form with $m_{\nu} \simeq m_D^2/M_{BL}$, where $m_D$
is the neutrino Dirac mass. 
In this case the low-energy effective theory
is the MSSM with unbroken $R$-parity, and contains besides the usual 
MSSM states, a triplet of Higgs scalars much lighter than the 
$B-L$ breaking scale.

A second option is to add non-renormalizable terms
to the Lagrangian of the minimal left-right supersymmetric model,
while retaining the minimal Higgs content
\cite{Martin:1992mq,AMRS,AMS}
The superpotential for this class of models 
can be written as
\begin{eqnarray}
W_{NR} &=& W_{min} +  {a_L \over 2 M} ({\rm Tr}\, \Delta_L \delta_L)^2
 +  {a_R \over 2 M} ({\rm Tr}\, \Delta_R \delta_R)^2+ {c \over M}{\rm Tr}\,
 \Delta_L \delta_L {\rm Tr}\, \Delta_R \delta_R \nonumber \\
& & + {b_L \over 2 M} {\rm Tr}\, \Delta_L^2 {\rm Tr}\, \delta_L^2 +
 {b_R \over 2 M} {\rm Tr}\, \Delta_R^2 {\rm Tr}\, \delta_R^2 + {1 \over M}
 \left[ d_1{\rm Tr}\, \Delta_L^2 {\rm Tr}\,  \delta_R^2 +
d_2 {\rm Tr}\, \delta_L^2 {\rm Tr}\, \Delta_R^2 \right]
  \nonumber \\
 & &
+ {\lambda_{ijkl}\over  M}{\rm Tr}\,i\tau_2 \Phi_i^T i\tau_2
\Phi_j{\rm Tr}\,i\tau_2 \Phi_k^T i\tau_2
\Phi_l + {\alpha_{ijL}\over M} {\rm Tr}\, \Delta_L  \delta_L
\Phi_i i\tau_2 \Phi_j^T i\tau_2   \nonumber\\
&&+  {\alpha_{ijR}\over M}
{\rm Tr}\, \Delta_R  \delta_R  i\tau_2 \Phi_i^T i\tau_2 \Phi_j 
+ {1 \over M}{\rm Tr}\,\tau_2 \Phi_i^T \tau_2 \Phi_j [\beta_{ijL}
{\rm Tr}\, \Delta_L
 \delta_L + \beta_{ijR} {\rm Tr}\, \Delta_R \delta_R ] \nonumber \\
& &+ 
{\eta_{ij} \over M } {\rm Tr}\,\Phi_i  \Delta_R i\tau_2
 \Phi_j^T i \tau_2 \delta_L
+
{\overline \eta_{ij} \over M } {\rm Tr}\,\Phi_i  \delta_R
i\tau_2 \Phi_j^T i\tau_2
 \Delta_L  \nonumber \\
& & + { k_{ql} \over M}Q^T i\tau_2 L\,Q^{cT} i\tau_2 L^c
 + {k_{qq} \over M}Q^T i\tau_2 Q\,Q^{cT} i\tau_2 Q^c
+ { k_{ll} \over M}L^T i\tau_2 L\,L^{cT} i\tau_2 L^c
\nonumber \\ & &+
{1\over M} [j_L Q^T i\tau_2 Q\,Q^T i\tau_2 L +  j_R
Q^{cT} i\tau_2 Q^c \,Q^{cT} i\tau_2 L^c].
\label{nonsuperpot}
\end{eqnarray}
It has been shown that the addition of non-renormalizable
terms suppressed by a high scale such as Planck mass,
$M \sim M_{Planck} \sim 10^{19}$ GeV, with the minimal field content
ensures the correct pattern of symmetry breaking
in the supersymmetric left-right model. 
In particular the scale of
parity breakdown is predicted to be in the intermediate
region $M_R \gsim 10^{10} - 10^{11}$ GeV, and $R$-parity remains
exact. This theory contains singly charged and
doubly charged Higgs scalars with a mass of order
$M_R^2/M_{Planck}$, which may be experimentally accessible.
However, what is different is the nature of see-saw mechanism.
Whereas in the renormalizable version the see-saw mechanism
takes its canonical form, in the non-renormalizable
case it takes a form similar to what occurs in the 
non-supersymmetric left-right models, with 
the neutrino mass depending  on the unknown 
parameters of the Higgs potential. This in general leads to
different neutrino mass spectra, which 
can be experimentally distinguished.

\section{The tree-level upper bound on the lightest Higgs mass}
\label{sec:higgs}

Given the fact that the Higgs sector of SLRM models contain a
large number of Higgs multiplets, and the VEVs of some of the Higgs
fields involve possibly large mass scales compared to the electroweak
and SUSY breaking scales, it is important to ask what is the mass 
of the lightest Higgs boson in these models.
The upper bound on the lightest Higgs boson mass
in the minimal model was derived in \cite{Huitu:1997rr} in the limit when
$\kappa_1',\kappa_2',\sigma_L,v_{\Delta_L},v_{\delta_L} \rightarrow 0$,
using the fact that for any Hermitean
matrix the smallest eigenvalue must be smaller than that of its
upper left corner $2 \times 2$ submatrix.
In the basis in which the first two indices correspond to
($\Phi^0_1$,$\chi^0_2$),  we find for matrix elements
$m_{11}^2,m_{22}^2,m_{12}^2$ (see Appendix \ref{sec:1})
\begin{eqnarray}
m_{11}^2&=& -m_{\Phi\chi }^2\frac{\kappa_2}{\kappa_1}
+\frac 12 (g_L^2+g_R^2)\kappa_1^2,\nonumber\\
m_{22}^2&=& -m_{\Phi\chi }^2\frac{\kappa_1}{\kappa_2}
+\frac 12 (g_L^2+g_R^2)\kappa_2^2,\nonumber\\
m_{12}^2&=& m_{\Phi\chi }^2
-\frac 12 (g_L^2+g_R^2)\kappa_1\kappa_2.
\end{eqnarray}
It follows that the upper bound on the lightest Higgs boson mass
in the minimal supersymmetric left-right model can be
written as  \cite{Huitu:1997rr}:
\begin{equation}
m_{h}^2\leq \frac 12 (g_L^2+g_R^2)(\kappa_1^2 +\kappa_2^2)\cos^2 
2\beta = \left(1+\frac{g_R^2}{g_L^2} \right) m_{W_L}^2 \cos^2 2 \beta
,
\label{eq:treeupper1}
\end{equation}
where $\tan\beta = \kappa_2/\kappa_1 $. The upper bound
(\ref{eq:treeupper1}) is not only independent of the supersymmetry
breaking parameters (as in the case of supersymmetric models based on
$SU(2)_L \times U(1)_Y$), but it is also independent of the $SU(2)_R$
breaking scale, which, {\em a priori}, can be large. The upper bound
is controlled by the weak scale vacuum expectation 
value,  $\kappa_1^2+\kappa_2^2$, and the
dimensionless gauge couplings ($g_L$ and $g_R$) only. Since the former
is essentially fixed by the electroweak scale, the gauge couplings
$g_L$ and $g_R$ determine the bound. 

We will see below that even when 
$\kappa_1',\kappa_2',\sigma_L,v_{\Delta_L},v_{\delta_L}$
are non-zero, the
upper bound on the lightest Higgs mass at the tree level 
does not depend on either 
the right-handed breaking scale or the SUSY breaking scale.
A general method to find an upper limit for the lightest Higgs mass  
in models based on $SU(2)_L\times U(1)_Y$ and maximally quartic potentials 
was presented in \cite{Comelli:1996xg}.
We will  apply this method to the case of SLRMs,
with possible nonrenormalizable terms, in an appropriate manner.

Consider a set of scalar fields $\Phi_j$ transforming under
$SU(2)_L$, and define a 
discrete transformation $P:\, \Phi_j\rightarrow
(-1)^{2T_j}\Phi_j$ of these fields, where $2T_j+1$ is the dimension of 
$SU(2)_L$ representation\footnote{One could as well choose any other
spontaneously broken $U(1)$ \cite{Comelli:1996xg}, but $P$
provides the best limit in the case of SLRM.}.
By setting  the $P$-even fields to their VEVs, a normalized field can be
defined in the direction of $SU(2)_L$ breaking,
$\phi=\frac 1{v_0}\sum_{odd} v_i\Phi_i$,
where $v_0^2=\sum_iv_i^2$, and all the fields orthogonal to $\phi$ have
zero VEVs.
Since the original lagrangian is invariant under the transformatiom $P$,
the potential must be invariant under $P$, so that the 
potential contains only  even powers of $\phi$,
\begin{eqnarray}
V(\phi )=V(0)-\frac 12 m^2\phi^2 +\frac 18 \lambda_\phi \phi^4 
+\frac 16 A\phi^6+\dots .
\end{eqnarray}
We take into account only the leading nonrenormalizable terms and thus
$\phi^6$ is the largest power in the potential $V(\phi)$.
If $\phi$ were the mass eigenstate, then by using the minimization condition 
the Higgs mass would be $\lambda_\phi v_0^2 + 4Av_0^4$.
In the general case, this expression provides an upper bound 
on the mass of the lightest Higgs boson,
\begin{eqnarray}
m_h^2\le \lambda_\phi v_0^2 + 4Av_0^4.
\end{eqnarray}

Since only the $SU(2)_L$ doublet fields are relevant for obtaining this
bound, adding extra singlets or triplets in the model has no effect
on the bound.
Thus, there are three separate cases to be considered for
deriving an upper bound on the mass of the lightest Higgs boson
in the models discussed in the last section, 
namely: (A) $R$-parity is broken (sneutrinos get VEVs),
(B) $R$-parity is conserved because there are additional triplets, and
(C) $R$-parity is conserved because there are nonrenormalizable terms.

Let us first consider the situation in the minimal model, case (A).
We define a new neutral scalar field in the direction of the breaking
in the $SU(2)_L$ doublet space:
\begin{equation}
\Phi^0 = \frac 1v \left( \kappa_1 \rm{Re} \,\Phi^0_1 +\kappa_1'
\rm{Re} \,\Phi^0_2+ \kappa_2' \rm{Re} \,\chi^0_1 + \kappa_2 \rm{Re}
\,\chi^0_2 + \sigma_L \rm{Re}  \,\tilde\nu \right) ,
\end{equation}
where
\begin{equation}
v^2 = \kappa_1^2+ \kappa_1'^2+\kappa_2^2+\kappa_2'^2+\sigma_L^2 .
\end{equation}
All the $SU(2)_L$ doublet fields which are orthogonal to $\Phi^0$ have
vanishing VEVs.
We calculate next the quartic term $(\Phi^0)^4$ in the potential.
It has contributions from both the F-terms
and the D-terms,
and we find that the upper bound on the mass of
the lightest Higgs boson in the minimal SLRM is
\begin{equation}
m_h^2 \leq \frac 1{2 v^2} \left[ g_L^2 (\omega_\kappa^2+\sigma_L^2)^2+
g_R^2 \omega_\kappa^4+g_{B-L}^2 \sigma_L^4 + 8 (h_{\Phi L}
\kappa_1'+h_{\chi L} \kappa_2 )^2 \sigma_L^2 + 8 h_{\Delta_L}^2
\sigma_L^4 \right] ,
\label{eq:treeupper2}
\end{equation}
where
\begin{equation}
\omega_\kappa^2 = \kappa_1^2-\kappa_2^2-\kappa_1'^2+\kappa_2'^2 .
\end{equation}
As is evident, the upper bound (\ref{eq:treeupper2}) is independent of
SUSY and right-handed breaking scales, and depends only on the
dimensionless gauge and Yukawa couplings, and vacuum expectation
values which are determined by the weak scale:
\begin{equation}
m_{W_L}^2 = \frac 12 g_L^2  \left(
\kappa_1^2+\kappa_2^2+\kappa_1'^2+\kappa_2'^2 +\sigma_L^2+ 2
v_{\Delta_L}^2+2 v_{\delta_L}^2 \right) 
+{\cal{O}}\left(\frac{\kappa^{'2} m_{W_L}^2}{m_{W_R}^2}
\right).
\end{equation}
The triplet VEVs $v_{\Delta_L}$ and $v_{\delta_L}$ must be small in order to
maintain $\rho \simeq  1$.
In the limit when $\kappa_1',\kappa_2',\sigma_L,v_{\Delta_L},
v_{\delta_L} \rightarrow 0$,
the bound (\ref{eq:treeupper2}) reduces to the upper bound 
(\ref{eq:treeupper1}).

It is obvious that the addition of extra triplets does not change 
this bound.
Thus, the bound for the case (B), the SLRM with additional triplets
to ensure that $R$-parity is not spontaneously broken,
can be obtained from
(\ref{eq:treeupper2}) by taking the limit $\sigma_L\rightarrow 0$.

The total number of nonrenormalizable terms in case (C) is rather 
large.
All the coefficients in nonrenormalizable terms are proportional to 
inverse powers of a large scale.
Thus the largest contribution comes from those terms which have the
smallest number of large scales and the largest number of 
potentially large VEVs from $SU(2)_R$ triplets.
We recall that the terms of the form $(\Phi^0)^4$ and $(\Phi^0)^6$
in the potential are needed to determine 
the contributions to the mass bound.
The nonrenormalizable terms have no effect on the contribution
from $D$-terms.
The leading terms in the superpotential which can give a 
$(\Phi^0)^4$ and $(\Phi^0)^6$ type F-term contribution are of the type
\begin{eqnarray}
W_{NR}=A{\rm Tr}(i\tau_2\Phi^Ti\tau_2\chi)\,{\rm Tr}(\Delta_R\delta_R)+
B{\rm Tr}(i\tau_2\Phi^Ti\tau_2\chi)^2,
\end{eqnarray}
i.e. one term with two and another with four bidoublet fields.
Here $A,\,B\sim 1/M_{\rm Planck}$.
With $SU(2)_L$ singlets fixed to their VEVs, the corresponding $F$-terms
are
\begin{eqnarray}
V_{NR}={\cal{O}}(ABv_{\Delta_R}v_{\delta_R})(\Phi^0)^4
+{\cal{O}}(B^2)(\Phi^0)^6.
\end{eqnarray}
The contribution to the Higgs mass bound from these nonrenormalizable
terms is  
\begin{equation}
{\cal{O}}(v_R^2/M_{\rm Planck}^2)\langle \Phi^0\rangle ^2 +
{\cal{O}}(1/M_{\rm Planck}^2)\langle \Phi^0\rangle ^4.
\end{equation}
If the VEV $v_R\sim
10^{10}$ GeV in these models,  
the contribution is numerically negligible.
Therefore the upper bound for this class of models is essentially the
same as in the case (B).

\section{Radiative corrections}
\label{sec:rad}

Since it is known that the radiative corrections to the lightest Higgs
mass are significant in the MSSM, as well as its extensions based on
the $SU(2)_L \times U(1)_Y$, it is important to
consider the radiative corrections
to the upper bound on the lightest Higgs boson mass obtained
in the  previous section.
In this section we discuss the one-loop radiative corrections to the
upper bound on the lightest Higgs mass in the minimal supersymmetric
left-right model, which was
obtained in last section. We shall use the method of one-loop
effective potential \cite{Coleman:1973tz} for the calculation of radiative
corrections, where the effective potential may be expressed as the sum
of the tree-level potential plus a correction coming from the sum of
one-loop diagrams with external lines having zero momenta,
\begin{equation}
V_{\rm{1-loop}} = V_{\rm{tree}} + \Delta V_1 ,
\end{equation}
where $ V_{\rm{tree}}$ is the tree level potential
(\ref{eq:scalarpotential}) evaluated at the appropriate running scale
$Q$, and $\Delta V_1$ is the one loop correction given by
\begin{equation}
\Delta V_1 = \frac 1{64 \pi^2} \sum_i (-1)^{2 J_i} (2 J_i+1) m_i^4
\left( \ln \frac{m_i^2}{Q^2} - \frac 3 2 \right) ,
\label{eq:deltav1}
\end{equation}
where $m_i$ is the field dependent mass eigenvalue of the $i$th
particle of spin $J_i$. The dominant contribution to
(\ref{eq:deltav1}) comes from top-stop ($t-\tilde t$) system. However,
under certain conditions the contribution of bottom-sbottom ($b-\tilde b$) 
can be nonneglible. We shall include both these contributions in our
calculations of the radiative corrections.

In order to evaluate the contributions of top-stop and bottom-sbottom
to (\ref{eq:deltav1}), we need the stop and sbottom mass matrices for
the SLRM. From (\ref{eq:fterms}), (\ref{eq:dterms}) and
(\ref{eq:softterms}), it is straightforward to calculate squark mass
matrices \cite{Huitu:1997iy}. Ignoring the interfamily mixing, the
part of the potential containing the stop and sbottom mass terms can
be written as
\begin{equation}
V_{squark}=\left( \begin{array}{cc} U_L^* & U_R^* \end{array} \right)
\tilde M_U \left( \begin{array} {c} U_L \\ U_R \end{array} \right) +
\left( \begin{array} {cc} D_L^* & D_R^* \end{array} \right) \tilde M_D
\left( \begin{array} {c} D_L \\ D_R \end{array} \right) ,
\end{equation}
where the mass matrix elements for the stop are
\begin{eqnarray}
(\tilde M_U)_{  U_L^*   U_L} &=&  \tilde m_Q^2+
m_u^2 +
\frac 14 g_L^2 \omega^2_\kappa  -
\frac 16 g_{B-L}^2 \omega^2_R  ,
\nonumber\\
(\tilde M_U)_{  U_R^*   U_L} &=&
h_{\Phi Q} A_{\Phi Q} \kappa '_1+h_{\chi Q} A_{\chi Q}\kappa_2 - 
\mu_1 (h_{\phi Q}\kappa '_2+h_{\chi Q}\kappa_1)-
2h_{\phi Q}\mu '_1 \kappa_1-
2 h_{\chi Q}\mu''_1\kappa '_2\nonumber\\
&&+
(h_{\phi L }h_{\phi Q } +h_{\chi L }h_{\chi Q }) \sigma_L\sigma_R \nonumber\\
&=&\left[(\tilde M_U)_{  U_L^* U_R }\right]^*,\nonumber\\
(\tilde M_U)_{  U_R^*   U_R} &=& m_{Q^c}^2+
m_u^2 +
\frac 14 g_R^2(\omega^2_\kappa -2\omega^2_R ) +
\frac 16 g_{B-L}^2 \omega^2_R  
,
\end{eqnarray}
while for the sbottom these are
\begin{eqnarray}
(\tilde M_D)_{  D_L^*   D_L} &=&
  m_Q^2+
m_d^2
-\frac 14 g_L^2 \omega^2_\kappa  -
\frac 16 g_{B-L}^2 \omega^2_R  
,\nonumber\\
(\tilde M_D)_{  D_R^*   D_L} &=&
-h_{\Phi Q} A_{\Phi Q} \kappa_1-h_{\chi Q} A_{\chi Q}\kappa '_2 + 
\mu_1 (h_{\phi Q}\kappa_2+h_{\chi Q}\kappa '_1) +
2h_{\phi Q}\mu '_1 \kappa '_1+
2 h_{\chi Q}\mu''_1\kappa_2
\nonumber\\
& = &\left[(\tilde M_D)_{  D_L^* D_R  }\right]^*,\nonumber\\
(\tilde M_D)_{ D_R^*  D_R  } &=& m_{Q^c}^2+
m_d^2
-\frac 14 g_R^2(\omega^2_\kappa -2\omega^2_R ) +
\frac 16 g_{B-L}^2 \omega^2_R  ,
\end{eqnarray}
where top and sbottom squared masses are given by $(h_{\chi Q}
\kappa_2)^2$ and $(h_{\Phi Q} \kappa_1)^2$, respectively, $m
^2_Q$, $m ^2_{Q^c}$, $A_{\Phi Q}$ and
$A_{\chi Q}$ are soft supersymmetry breaking parameters (see eq.
(\ref{eq:softterms})), and
\begin{equation}
\omega^2_R=v_{\Delta_R}^2-v_{\delta_R}^2-\frac 12\sigma_R^2, \;
\omega^2_\kappa = \kappa_1^2+{\kappa '_2}^2-\kappa_2^2-{\kappa '_1}^2.
\end{equation}
In order that $SU(3)_C \times U(1)_{\rm{em}}$ is unbroken, none of
the physical squared masses of squarks can be negative. Necessarily
then all the diagonal elements of the squark mass matrices should be
non-negative. Combining the diagonal elements of the stop and sbottom
mass matrices leads to the inequality
\begin{equation}
m_Q^2+m_{Q^c}^2 \geq |\frac 12 g_R^2\omega^2_R|=
\frac 12g_R^2|v_{\Delta_R}^2-
v_{\delta_R}^2 -\frac 12 \sigma_R^2 |.
\label{eq:susyineq}
\end{equation}
where we have ignored terms which are of the order of the weak scale
or less.

The eigenvalues $m^2_{\tilde t _{1,2}},m^2_{\tilde b _{1,2}}$ of the
stop and sbottom mass squared matrices are given by ($m^2_{\tilde t
_1} > m^2_{\tilde t _2}, m^2_{\tilde b _1} > m^2_{\tilde b _2}$)
\begin{eqnarray}
m^2_{\tilde t_{1,2}} = m^2_{\tilde t} \pm \Delta^2_{\tilde t} ,
\nonumber \\
m^2_{\tilde b_{1,2}} = m^2_{\tilde b} \pm \Delta^2_{\tilde b} ,
\label{eq:squark1}
\end{eqnarray}
where
\begin{eqnarray}
m^2_{\tilde t}& = &\frac 12 \left[  m ^2_Q+ m ^2_{Q^c} + 2
m^2_t + \frac 14 (g_L^2+g_R^2) \omega_\kappa^2-\frac 12 g_R^2
\omega_R^2 \right] , \nonumber \\ 
\Delta^2_{\tilde t} &= &\frac 12
\left\{ \left[  m ^2_Q- m ^2_{Q^c} + \frac 14
(g_L^2-g_R^2) \omega_\kappa^2 + \frac 12 g_R^2 \omega_R^2- \frac 13
g_{B-L}^2 \omega_R^2 \right] ^2 \right. \nonumber \\ && \left. + 4
\left[ h_{\chi Q} A_t \kappa_2 - h_{\chi Q}\tilde\mu_t
\kappa_1 \right]^2 \right\} ^{\frac 12} ,
\label{eq:squark2}
\end{eqnarray}
and
\begin{eqnarray}
m^2_{\tilde b} &= &\frac 12 \left[  m ^2_Q+ m ^2_{Q^c} + 2
m^2_b - \frac 14 (g_L^2+g_R^2) \omega_\kappa^2+\frac 12 g_R^2
\omega_R^2 \right] , \nonumber \\ 
\Delta^2_{\tilde b} &= &\frac 12
\left\{ \left[  m ^2_Q- m ^2_{Q^c} - \frac 14
(g_L^2-g_R^2) \omega_\kappa^2 - \frac 12 g_R^2 \omega_R^2- \frac 13
g_{B-L}^2 \omega_R^2 \right] ^2 \right. \nonumber \\ && \left. + 4
\left[ h_{\Phi Q} A_b \kappa_1 -  h_{\Phi Q}\tilde\mu_b
\kappa_2 \right]^2 \right\} ^{\frac 12} .
\label{eq:squark3}
\end{eqnarray}
Here we have defined
\begin{eqnarray}
&&\tilde\mu_t\equiv \mu_1+2 \mu_1' \frac{m_b}{m_t} \tan \beta,\;\;
\tilde\mu_b\equiv \mu_1+2 \mu_1''\frac{m_t}{m_b} \cot \beta,\;\;
A_t\equiv A_{\chi Q},\;\; A_b\equiv A_{\phi Q}.
\end{eqnarray}
Using eqs.\ (\ref{eq:squark1}), (\ref{eq:squark2}) and
(\ref{eq:squark3}) in (\ref{eq:deltav1}), we have calculated the
radiatively-corrected expressions for the matrix elements of the upper
left corner $2 \times 2$ submatrix of the $10 \times 10$ CP-even Higgs
mass matrix. After imposing the appropriate one-loop minimization
conditions, we find the following form for the radiatively corrected
upper left corner $2 \times 2$ submatrix of CP-even Higgs mass matrix:
\begin{eqnarray}
\left( \begin{array}{cc} \frac 12 (g_L^2+g_R^2) \kappa_1^2 & - \frac
12 (g_L^2+g_R^2) \kappa_1 \kappa_2 \\  - \frac 12 (g_L^2+g_R^2)
\kappa_1 \kappa_2 &  \frac 12 (g_L^2+g_R^2) \kappa_2^2 \end{array}
\right) 
+ \left( \begin{array}{cc} \tan \beta & -1 \\ -1
& \cot \beta \end{array} \right)  \left( \frac  \Delta 2 \right) +
\frac{3 g_L^2}{16 \pi^2 m_{W_L}^2} \left( \begin{array}{cc} \Delta_{11}
& \Delta_{12} \\ \Delta_{12} & \Delta_{22} \end{array} \right) ,\nonumber\\
\label{eq:squarkmass}
\end{eqnarray}
where
\begin{eqnarray}
\label{eq:radelements}
\Delta &=&  \left[ -2 m^2_{\Phi \chi} + \frac 3 {32 \pi^2} \left(
\frac{g_L^2}{\sin^2 \beta} \frac{m_t^2}{m_W^2} \right) A_t 
\tilde\mu_t 
\frac{f(m_{\tilde t_1}^2)-f(m_{\tilde t_2}^2)}
{m_{\tilde t_1}^2-m_{\tilde t_2}^2}
\right. 
\nonumber\\ &&
\left. + \frac 3 {32 \pi^2} \left( \frac{g_L^2}{\cos ^2 \beta}
\frac{m_b^2}{m_W^2} \right) A_b \tilde\mu_b 
\frac{f(m_{\tilde b_1}^2)-f(m_{\tilde b_2}^2)}
{m_{\tilde b_1}^2-m_{\tilde b_2}^2}
 \right] ,
\label{eq:radelementsa}
\end{eqnarray}
\begin{eqnarray}
&&f(x^2)=2x^2(\ln(x^2/Q^2)-1),
\label{def:f}  
\end{eqnarray}
and
\begin{eqnarray}
\Delta_{11} &= & \frac{m_b^4}{\cos^2 \beta} \left[ \ln \left( \frac{
m^2_{\tilde b_1} m^2_{\tilde b_2}}{m_b^4} \right) + \frac{2 A_b \left( A_b -
\tilde\mu_b \tan \beta
\right) }{m^2_{\tilde b _1}-m^2_{\tilde b _2}} \ln \left(
\frac{m^2_{\tilde b_1}}{m^2_{\tilde b _2}} \right) \right] \nonumber
\\ && + \frac{m_b^4}{\cos^2 \beta} \frac{A_b^2 \left( A_b - \tilde
\mu_b \tan \beta \right)
^2 }{\left( m^2_{\tilde b _1} - m^2_{\tilde b _2} \right) ^2} g (
m^2_{\tilde b _1} , m^2_{\tilde b _2} ) 
 + \frac{m_t^4}{\sin^2 \beta} \frac{\left( A_t - 
\tilde\mu_t \cot \beta \right)
^2  \tilde\mu_t  ^2 }{
(m^2_{\tilde t _1} - m^2_{\tilde t _2})^2 } g (
m^2_{\tilde t _1} , m^2_{\tilde t _2} ) ,\nonumber\\
\label{eq:radelementsb}
\end{eqnarray}
\begin{eqnarray}
\Delta_{22}& = & \frac{m_t^4}{\sin^2 \beta} \left[ \ln \left(
\frac{m^2_{\tilde t_1 } m^2_{\tilde t _2}}{m_t^4} \right)+ \frac{2 A_t
\left( A_t- \tilde\mu_t
\cot \beta \right) }{m^2_{\tilde t _1}-m^2_{\tilde t _2}} \ln \left(
\frac{m^2_{\tilde t _1}}{m^2_{\tilde t _2}} \right) \right] \nonumber
\\
&& + \frac{m_t^4}{\sin^2 \beta} \frac{A_t^2 \left( A_t - \tilde \mu_t
 \cot \beta \right)
^2}{\left( m^2_{\tilde t _1}-m^2_{\tilde t _2} \right) ^2 } g(
m^2_{\tilde t _1},m^2_{\tilde t _2}  ) 
 + \frac{m_b^4}{\cos ^2 \beta} \frac{ \left( A_b - \tilde \mu_b
 \tan \beta \right) ^2 \tilde\mu_b  ^2 }{
\left( m^2_{\tilde b _1}- m^2_{\tilde b _2} \right) ^2 } g(m^2_{\tilde
b _1}, m^2_{\tilde b _2}) ,\nonumber\\
\label{eq:radelementsc}
\end{eqnarray}
\begin{eqnarray}
\Delta_{12} &= & \frac{m_t^4}{\sin^2 \beta} \frac{ \left( A_t - \tilde
\mu_t  \cot \beta \right)
\left( -\tilde\mu_t \right) }{m^2_{\tilde
t _1}-m^2_{\tilde t_2}}  \times \left[ \ln \left(
\frac{m^2_{\tilde t_1}}{m^2_{\tilde t _2}} \right) + \frac{ A_t \left(
A_t - \tilde\mu_t \cot
\beta \right) }{m^2_{\tilde t _1}-m^2_{\tilde t _2}} g( m^2_{\tilde t
_1}, m^2_{\tilde t _2} ) \right] \nonumber \\ 
&& + \frac{m_b^4}{\cos^2
\beta} \frac{ \left( A_b - \tilde\mu_b  
\tan \beta \right) \left( -\tilde\mu_b \right) }
{m^2_{\tilde b _1}-m^2_{\tilde b
_2}}  \times \left[ \ln \left( \frac{m^2_{\tilde b
_1}}{m^2_{\tilde b _2}} \right) + \frac{A_b \left( A_b- \tilde\mu_b
 \tan \beta \right)
}{m^2_{\tilde b _1}-m^2_{\tilde b _2}} g(m^2_{\tilde b _1},m^2_{\tilde
b _2} ) \right] ,\nonumber\\
\label{eq:radelementsd}
\end{eqnarray}
with
\begin{equation}
g(m_1^2,m_2^2) = 2- \frac{m_1^2 + m_2^2}{m_1^2-m_2^2} \ln \left(
\frac{m_1^2}{m_2^2} \right)  .
\label{eq:radelementse}
\end{equation}
We have neglected D-terms in the squark masses, because these are
small, and, since we are including only the quark-squark
contributions to $\Delta V_1$, in order to gain approximate
independence of the renormalization scale $Q$ (see also the inequality
(\ref{eq:susyineq})).

Using eqs.\ (\ref{eq:squarkmass}) and (\ref{eq:radelements}),
we obtain the one-loop radiatively corrected upper bound on the
lightest Higgs boson mass in the SLRM:
\begin{eqnarray}
m_h^2 &\leq& \frac 12 \left[ (g_L^2+g_R^2) \left( \kappa_1^2+\kappa_2^2
\right) \cos^2 2 \beta + \frac{3 g_L^2}{8 \pi^2 m^2_{W_L}} \left(
\Delta_{11} \cos^2 \beta + \Delta_{22} \sin^2 \beta + \Delta_{12} \sin
2 \beta \right) \right]\nonumber\\
\label{eq:upperbound1}
\end{eqnarray}
For $\tan\beta \lsim 20 $,
one can neglect the b-quark
contribution in the radiative corrections. Then, in 
the approximation \cite{Carena:1995bx,Casas:1995us}
\begin{equation}
|m_{\tilde t_1}^2 - m_{\tilde t_2}^2|
\ll |m_{\tilde t_1}^2 + m_{\tilde t_2}^2|,
\end{equation}
the upper bound (\ref{eq:upperbound1}) on the 
lightest Higgs mass reduces to
\begin{eqnarray}
m_h^2 &\leq& \frac 12 \left[ (g_L^2+g_R^2) \left( \kappa_1^2+\kappa_2^2
\right) \cos^2 2 \beta\right.\nonumber\\ 
&&\left. + \frac{3 g_L^2 m_t^4}{8 \pi^2 m^2_{W_L}} \left(
\ln (\frac{m_{\tilde t_1}^2 m_{\tilde t_2}^2}{m_t^4}) 
+2 \frac{\tilde A_t^2}{M_s^2} (1 - \frac{\tilde A_t^2}{12M_s^2}) - 
8 \frac{\mu_1''^4}{3M_s^4} \right ) \right ]
\label{eq:upperbound2}
\end{eqnarray}
where $\tilde A_t = A_t - \mu_1 \cot\beta$, and $M_s$ is the
supersymmetry breaking scale 
($2M_s^2 = m_{\tilde t_1}^2 + m_{\tilde t_2}^2$).
In this limit the upper bound eq.\ (\ref{eq:upperbound2}) 
on the lightest Higgs mass in 
the supersymmetric left-right model 
differs in form from the 
corresponding MSSM upper bound only because of
$\mu_1''$ being nonzero. 
The upper bound is maximised by
\begin{equation}
|\tilde A_t| = (\sqrt 6) M_s
\end{equation}
for a given value of $\mu_1''$. 

\begin{figure}[t]
\leavevmode
\begin{center}
\mbox{\epsfxsize=8.truecm\epsfysize=8.truecm\epsffile{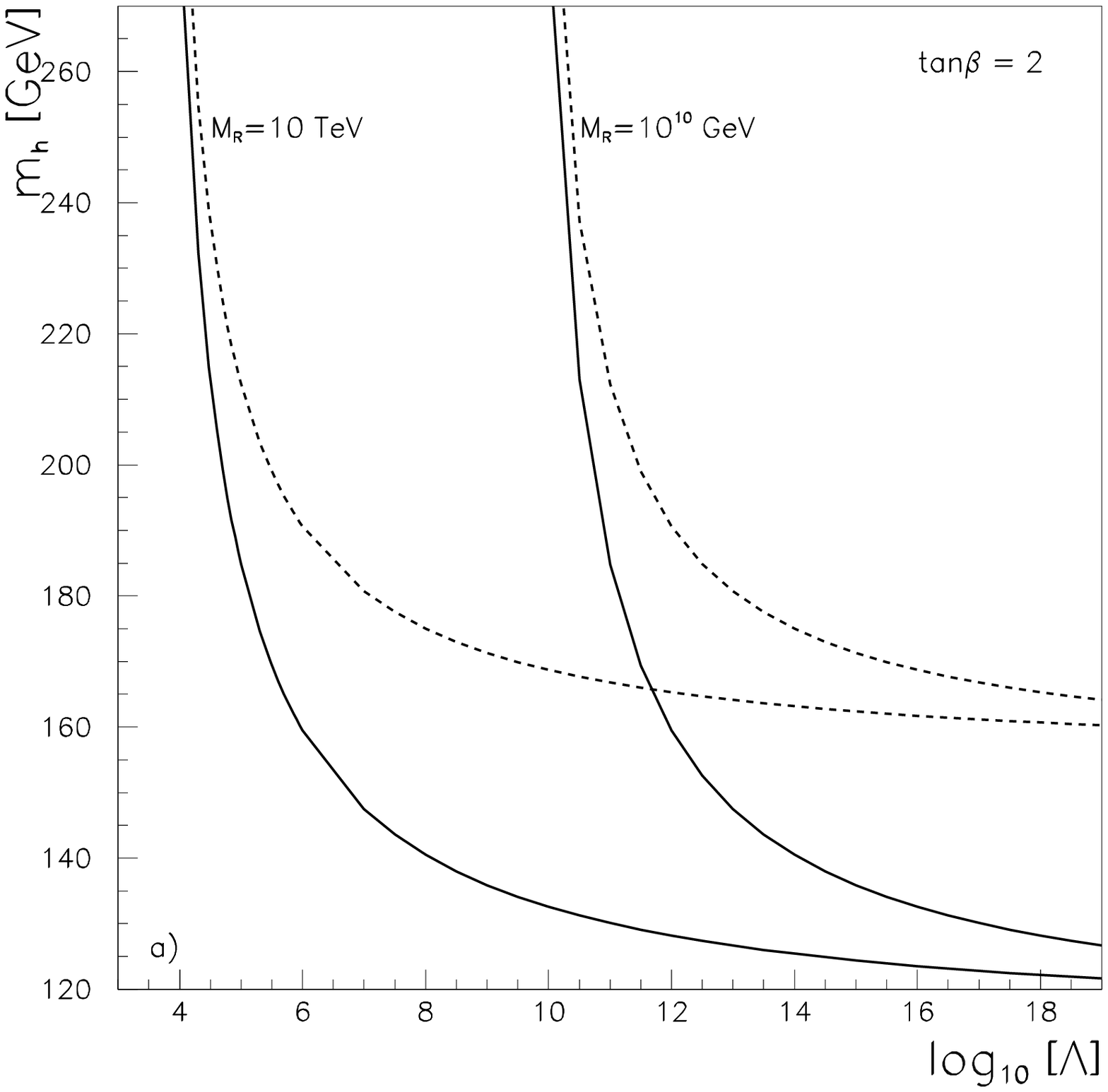}
\epsfxsize=8.truecm\epsfysize=8.truecm\epsffile{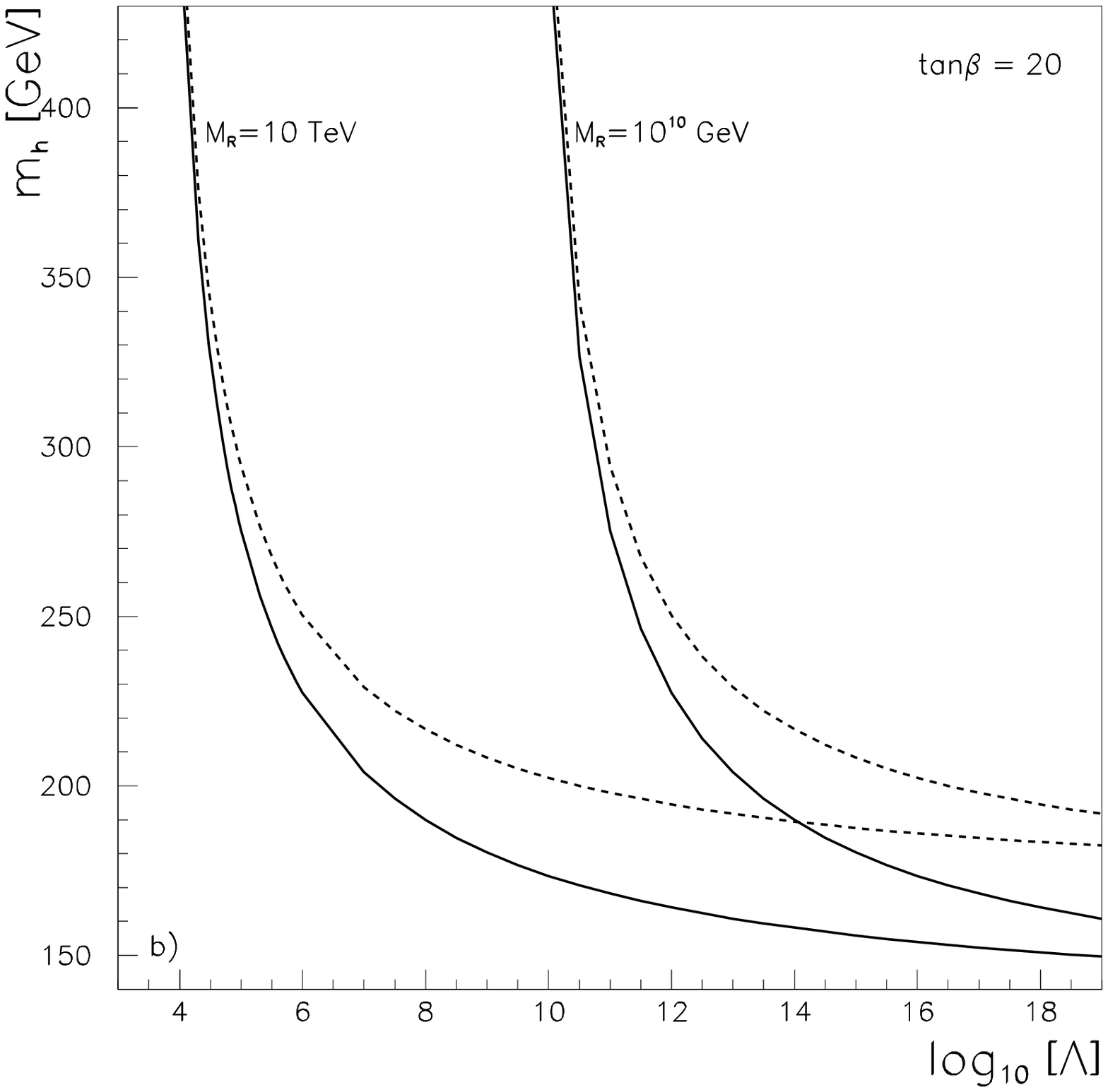}}
\end{center}
\caption{\label{mh1}The upper bound on the radiatively corrected mass 
of the lightest neutral Higgs boson for two different values of $\tan\beta$.
The right-handed scale $M_R$ is indicated in the Figure.
The bi- and trilinear soft supersymmetry breaking parameters are  1 TeV
(solid line) and 10 TeV (dashed line).
Supersymmetric Higgs mixing parameters are assumed to vanish, and
$m_{top}=175$ GeV.}
\end{figure}

The radiatively corrected upper bound (\ref{eq:upperbound1}) on 
the mass of the lightest Higgs boson  is plotted in Fig.\ref{mh1} 
as a function of the large scale $\Lambda $ up to
which the supersymmetric left-right model remains
perturbative.  
The upper bound comes from the requirement that 
all the gauge couplings of the SLRM
remain perturbative below the scale $\Lambda $.
We have taken into account the dominant radiative corrections 
coming from the quark and squark loops in our 
calculations.
In Fig.\ref{mh1} we have taken two values of
$\tan\beta=2$ and  $\tan\beta=20$.
In the figure the upper bound is shown for two different values of
the $SU(2)_R$ breaking
scale, $M_R=10$ TeV and $M_R=10^{10}$ GeV, respectively,
and for two values of soft supersymmetry breaking
mass parameter, $M_s=1$ TeV and $M_s=10$ TeV.
It is seen from this figure that if the difference between the $SU(2)_R$
breaking scale and the large scale $\Lambda$ 
is more than two orders of
magnitude, the radiatively corrected upper bound on the mass
of the lightest Higgs boson remains below 250 GeV.
For large values of $\Lambda $ the upper bound is below 200 GeV.
The upper bound increases with increasing $M_R$ and with increasing
soft supersymmetry breaking parameters.
It is considerably larger than the corresponding upper
bound in the MSSM.

\section{Couplings of the lightest neutral Higgs to fermions and the 
decoupling limit}
\label{sec:decoup}

In order to study the phenomenology of the lightest Higgs boson
in the SLRM, we must obtain  its couplings to the fermions and see
how these differ from the corresponding MSSM couplings.
The major difference between the two models arises 
due to the triplet Higgs couplings to leptons.  As discussed
earlier, the lightest neutral Higgs is composed mainly of the
bidoublet fields, and we do not expect the triplet couplings to have a
large effect on the lightest neutral Higgs branching ratios.  Another
difference in the Higgs couplings in the SLRM and the MSSM 
arises because of the mixing between the 
Higgs(ino) and lepton sectors in models where $R$-parity is
spontaneously broken.
Since the lightest chargino
contains gaugino and higgsino admixture, one might expect that 
in some region of the parameter space the couplings
are significantly different from the  couplings
in the MSSM. We will take the
lightest chargino to correspond to the $\tau $-lepton. It will be
denoted by $\tau$ in the following.  Thus, it is
important to study 
in detail the couplings of the lightest Higgs boson to the $\tau$.

\begin{figure}[t]
\leavevmode
\begin{center}
\mbox{\epsfxsize=8.truecm\epsfysize=8.truecm\epsffile{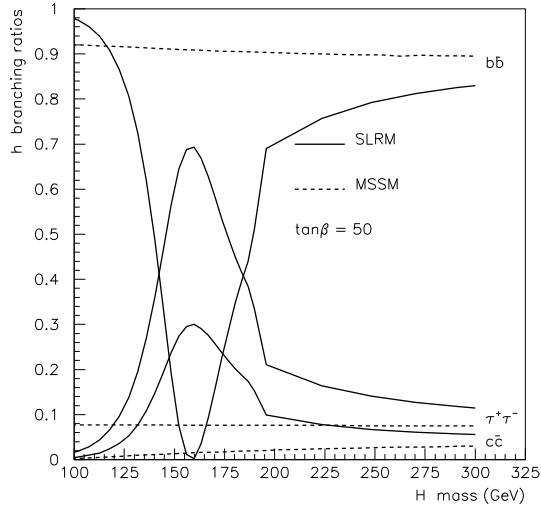}}
\end{center}
\caption{\label{tau2}  The branching ratio of the lightest 
neutral CP even Higgs boson $h$ to fermion pairs in SLRM (solid curves)
as a function of the second lightest Higgs $H$ mass.
The tree level mass of the lightest Higgs is $\sim$78 GeV and
$\tan\beta=50$.
The dotted curves correspond to MSSM.}
\end{figure}

In Figure \ref{tau2} we have plotted the branching ratio of
the lightest neutral Higgs boson to $b\bar b$, $c\bar c$, and
$\tau^+\tau^-$ pair both in the MSSM and in the SLRM for large
$\tan\beta$ and a tree-level mass $m_{h}\sim 78$ GeV as a function of
the second lightest neutral Higgs boson mass.  We note that for the
second lightest neutral Higgs with mass $\sim 160$ GeV, the branching
ratio to bottom pair almost vanishes while the branching ratio to a
$\tau$ pair is enhanced.  This behaviour occurs in a transition region
where the composition of the light Higgs bosons change rapidly,
and the lightest Higgs coupling to $b\bar b$ becomes negligible.  
Similar phenomenon occurs for small $\tan\beta$.

It is well known that in the left-right symmetric models problems with
FCNC are expected if several light Higgs bosons exist \cite{EGN}.
In the supersymmetric case, the problem could be expected to be more
severe, since the potential is more constrained.
Furthermore, in the four Higgs-doublet model \cite{MR}, 
which is also the number of doublets in the supersymmetric left-right 
model, it was found
that the potential is always real, thus preventing spontaneous 
violation of CP.
Strong limits on the FCNC Higgs were obtained in \cite{pospelov}
assuming that the only source for CP violation is the complex phase in
the Kobayashi-Maskawa matrix and that the model is manifestly
left-right symmetric.
If these conditions are relaxed, for example by defining the symmetry
transformation by Eq. (4), a weakened lower limit for second 
lightest neutral Higgs is obtained from the neutral meson mass 
difference\footnote{When R-parity is broken, the
ground state can violate CP via the soft breaking terms involving sneutrinos
\cite{mohapatra}.}.
The limit is less severe for large values of
$\tan\beta$, but remains of the same order than in the 
nonsupersymmetric left-right model.
Taking into account the uncertainties in calculating the mass
difference, $m_{H_{FCNC}}\gsim {\cal{O}}(1$ TeV). 
Thus the relevant limit to discuss is the one in which all the neutral Higgs
bosons, except the lightest one, are heavy.

In order to consider in detail the neutral Higgs couplings to the
$\tau$-leptons, we'll need the interaction Lagrangian.
We have seen that at the tree level the lightest Higgs can be written
as
\begin{eqnarray}
h=\frac{1}{v}\sum_k\left(1+{\cal{O}}\left(
\frac{m_h^2}{m_{H_2}^2}\right)\right) \langle\phi_k\rangle \phi_k +
\sum_k{\cal{O}}\left(
\frac{m_h}{m_{H_2}}\right)\psi_k ,
\end{eqnarray}
where $\phi_k$ are the scalar bidoublet fields and $\psi_k$ are all
the other fields.
In the decoupling limit $m_{H_2}\gg m_h$, the relevant 
Higgs-fermion interaction 
Lagrangian can be written as 
\begin{eqnarray}
L=-\frac 1v\psi_i^+ v_ky_{ijk}\psi_j^-h +h.c.,
\end{eqnarray}
where
$\psi^{-T}=(-i\lambda_L^-,-i\lambda_R^-,\phi_2^-,\chi_2^-,\Delta_R^-,
\tau )^T$ and 
$\psi^{+T}=(-i\lambda_L^+,-i\lambda_R^+,\phi_1^+,\chi_1^+,\delta_R^+,
\tau^c )^T$.
The mass eigenstates $\chi^\pm$ are found by diagonalizing the mass
matrices by unitary matrices $U,\,V$ 
\cite{Haber:1985rc},
\begin{eqnarray}
\chi^+=V\psi^+,\;\;\; \chi^-=U\psi^-.
\end{eqnarray}
The diagonal chargino mass matrix is 
\begin{eqnarray}
M_{\chi^\pm}^2=VX^\dagger XV^\dagger=U^*XX^\dagger U^T.
\end{eqnarray}
We decompose\footnote{For a more general treatment, see \cite{kai}.}
the mass matrix $X$ and the diagonalizing matrices $V$
and $U$ as
\begin{eqnarray}
X=X_0+\epsilon X_1,\;\;\; V=(1+\epsilon Y)V_0,\;\;\;
U=(1+\epsilon Z)U_0 ,
\end{eqnarray}
where the contribution of the terms proportional to $\epsilon $ in
determining the $\tau $ lepton eigenstate is small.
One can solve for the matrix elements to the order $O(\epsilon^2)$,
\begin{eqnarray}
Y_{jk}&=&\left\{\begin{array}{l}0,\;j=k,\\
\frac{[V_0(X_0^\dagger X_1+X_1X_0)V_0^\dagger]_{jk}}
{(M_0^2)_{jj}-(M_0^2)_{kk}},
\;j\neq k \end{array},\right . \\
Z_{jk}&=&\left\{\begin{array}{l}0,\;j=k,\\
\frac{[U_0^*(X_0X_1^\dagger+X_1X_0^\dagger )U_0^T]_{jk}}
{(M_0^2)_{jj}-(M_0^2)_{kk}},
\;j\neq k \end{array},\right . 
\end{eqnarray}
where we have denoted $M_0^2\equiv U_0^*X_0X_0^\dagger U_0^T$.

In terms of the mass eigenstates, the interaction Lagrangian can be written as
\begin{eqnarray}
L_{int}=-\frac 1v(\chi^{-T}U^*)_iv_ky_{ijk}(V^\dagger\chi^+)_j h +h.c.
\end{eqnarray}
Here we denote $C\equiv \frac 1v U^*v_ky_{ijk}V^\dagger $, where $C$
is the coupling matrix for charginos.
Thus the diagonal couplings to $\tau$ are given by
\begin{eqnarray}
C_{11}\sim\frac 1v \sum_{jk}U_{01j} X_{jk}V_{01k}^*.
\end{eqnarray}
The chargino mass matrix can be written as 
$X=X_0+\epsilon X_1$, where 
\begin{eqnarray}
X_0=\left(\begin{array}{cccccc}
m_L&0&0&0&0&0\\
0&m_R&0&0&\sqrt{2}g_R v_{\delta_R} &g_R\sigma_R\\
0&0&0&\mu_1&0&0\\
0&0&\mu_1&0&0&0\\
0&-\sqrt{2} g_Rv_{\Delta_R} &0&0&\mu_2&-\sqrt{2}h_\Delta\sigma_R\\
0&0&-h_{\Phi L}\sigma_R&-h_{\chi L}\sigma_R&0&0
\end{array}\right)
\end{eqnarray}
and
\begin{eqnarray}
\epsilon X_1=\left(\begin{array}{cccccc}
0&0&0&g_L\kappa_2&0&0\\
0&0&-g_R\kappa_1&0&0&0\\
g_L\kappa_1&0&0&0&0&0\\
0&-g_R\kappa_2&0&0&0&0\\
0&0&0&0&0&0\\
0&0&0&0&0&-h_{\Phi L}\kappa_1
\end{array}\right).
\end{eqnarray}
We have here assumed for simplicity that $\mu_1'=\mu_1''=0$, and
$\sigma_L=v_{\delta_L}=v_{\Delta_L}=0$.
The matrix $X_0$ has one zero eigenvalue corresponding to the 
$\tau $ mass, $(M_0)^2_{11}$=0.
The $\tau$ mass is 
\begin{eqnarray}
m_\tau &=&(U^*XV^\dagger)_{11}\nonumber\\
&=&(Z_1^*U_0^*X_0V_0^\dagger +U_0^*X_0V_0^\dagger
Y_1+U_0^*X_1V_0^\dagger)_{11}.
\end{eqnarray}
For the SM one has $v C_{11}=(U_0^*X_1V_0^\dagger )_{11}=m_\tau  $.
Thus, in the decoupling limit, 
we obtain the following result for the ratio of the 
Yukawa couplings $y_{SM}$ and $y_{SLRM}$:
\begin{eqnarray}
\frac{y_{SLRM}}{y_{SM}}=1+\frac{(Z_1^*M_0 -M_0 Y_1)_{11}}
{(U_0^*X_1V_0^\dagger)_{11}}=1,
\end{eqnarray}
since the matrix $M_0$ is diagonal.
Even if the $\tau$'s contained large fraction of gauginos or higgsinos,
the couplings responsible for Higgs decays to the charged 
leptons or quarks are the same as in the Standard Model, since the 
physically 
relevant parameter region is close to the decoupling limit. 

Contrary to the MSSM, one might have the lightest Higgs decaying to
a $\tau$-lepton and a heavier chargino.
Whether this decay mode is kinematically possible is much more model
dependent than the decay to leptons,  and we cannot say anything general
about it.

\section{The lightest doubly charged Higgs boson}
\label{sec:double}

We have seen that the neutral Higgs sector contains one relatively light
Higgs boson, which however may be heavier as compared to the 
lightest Higgs in the MSSM,  and which
has MSSM like couplings to the fermions of the model.
Thus,  one cannot tell from the properties of the lightest Higgs boson,
about  the nature of the model.
If a light Higgs is found before any supersymmetric particles
are observed, one
would not even know whether it is the Higgs boson of a
supersymmetric model.

Altogether there are four doubly charged Higgs bosons in the SLRM, of which
two are right-handed and two left-handed.
The mass matrices of the left-handed triplets depend on the soft
SUSY breaking parameters,  left-triplet VEVs, and the  
weak scale.
Thus,  their masses are expected to be of the same order as the soft terms.

The mass matrix for the right-handed scalars depends on the
right-triplet
VEV instead of the left-triplet VEV.
Nevertheless,
it was noticed in \cite{Huitu:1995zm} that in the SLRM with broken R-parity
the lightest doubly charged scalar tends to be light.
Also, it was shown in \cite{CM58} that in the nonrenormalizable case
it is possible to have light doubly charged Higgs bosons.
On the other hand, in the nonsupersymmetric left-right model all the
doubly charged scalars tend to have a mass of the order of the 
right-handed scale \cite{GGMKO}.
This is also true in the SLRM with enlarged particle content
\cite{AMRS}.
Thus a light doubly charged Higgs would be a strong indication of a
supersymmetric left-right model with minimal particle content.

\begin{figure}[t]
\leavevmode
\begin{center}
\mbox{\epsfxsize=8.truecm\epsfysize=8.truecm\epsffile{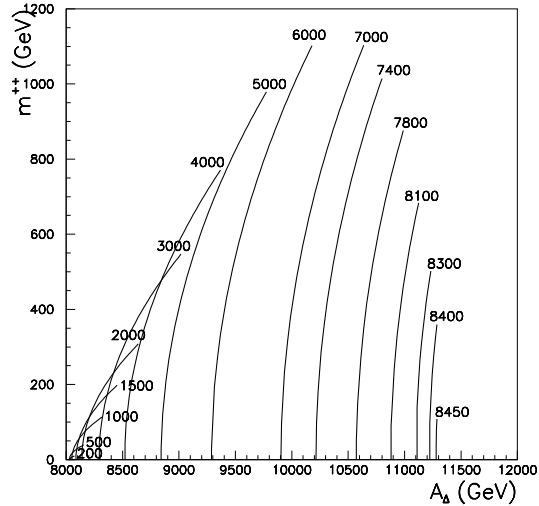}}
\end{center}
\caption{\label{mpp_ad}The mass $m^{++}$ of the lightest doubly charged Higgs
boson  as a function of the soft trilinear coupling $A_\Delta$
for different values of the right-handed sneutrino VEVs $\sigma_R$.
We have varied $\sigma_R$ in its allowed range of 100 GeV to 8.45
TeV as indicated in the figure.}
\end{figure}

In the case of broken R-parity, the mass matrix of the right-handed
doubly charged scalars can be obtained from the potential in appendix A.
The allowed parameter space is strongly constrained by demanding that
all the eigenvalues of the mass squared scalar matrices remain
positive.
In order to find the relevant parameter region, 
we have studied the bounds on some of the masses in the model.  
Although these may not be important as actual bounds, they
restrict the parameter space.
{}In the following we shall assume that $\tan\beta > 1$.
{}From the bound obtained from submatrix $(\Phi_2^{0r},\chi_1^{0r})$, we
find the following  constraint
\begin{eqnarray}
-\frac 12 g_L^2(\kappa_2^2-\kappa_1^2)-\frac 12 g_R^2D
-(h_{\chi L}^2\kappa_2^2-h_{\phi L}^2\kappa_1^2)\sigma_R^2\geq 0,
\label{constr1}
\end{eqnarray}
where
$D=2v_{\Delta_R}^2-2v_{\delta_R}^2-\sigma_R^2+\kappa_2^2-\kappa_1^2$.
If the terms with bidoublet lepton Yukawa couplings are ignored, and
$\tan\beta> 1$, Eq. (\ref{constr1}) indicates that $D<0$ and thus 
$2v_{\delta_R}^2+\sigma_R^2>2v_{\delta_R}^2$.
Other useful constraints follow 
from $(\Delta_R^{--},\delta_R^{++})$ and
$(\Delta_R^{0i},\delta_R^{0i})$ mass matrices,
\begin{eqnarray}
(A_\Delta v_{\Delta_R}-4h_{\Delta_R}^2v_{\Delta_R}^2+h_{\Delta_R}\mu_{2r}
v_{\delta_R})\sigma_R^2+g_R^2D(v_{\delta_R}^2-v_{\Delta_R}^2)&\geq& 0,
\nonumber\\
A_\Delta v_{\Delta_R}+h_{\Delta_R}\mu_{2r}v_{\delta_R}&\geq & 0.
\end{eqnarray}

We have studied in Figure \ref{mpp_ad} an example with the soft masses
and right-handed breaking scale, as well as the $\mu_{2R}$ parameter, 
of the order of 10 TeV.
The maximum Majorana Yukawa coupling allowed by positivity of the mass
eigenvalues in this case is 
$h_\Delta\sim 0.4$.
For $h_\Delta = 0.4$ we have plotted the allowed doubly charged Higgs
mass $m^{++}$ as a function of $A_\Delta $ for fixed 
$\sigma_R= 100$ GeV$,\dots,8.45$ TeV.
It is seen that relatively narrow bands of $A_\Delta$ and $\sigma_R$
are allowed.
Even in the maximal case the mass of the
doubly charged scalar $m^{++}\sim 1 $ TeV and from the figure 
we see that the lightest of the doubly charged scalars 
can be as light or even lighter than the lightest neutral Higgs boson.
The mass of the dangerous flavour changing Higgs boson is 
$\sim \frac{1}{\sqrt{2}} g_R \sqrt{D}$.

The mass of the lightest doubly charged Higgs, $m_{H_1^{++}}$, in the case of 
nonrenormalizable couplings has been considered in \cite{CM58,DM,AMRS}.
A general statement in these works is that the mass is $m_{H_1^{++}}
\sim v_R^2/M $, where $M$ is the scale of the
nonrenormalizable terms.
Relevant constraints on the parameters follow again from the
positivity of the mass bounds.
{}From the submatrix $(\Delta_R^-,\delta_R^+)$, a bound on $m_{H_1^+}$
can be obtained 
and we get
\begin{eqnarray}
0
&<&\frac 12 g_R^2 (v_{\delta_R}^2-v_{\Delta_R}^2)
(\kappa_2^2-\kappa_1^2).
\end{eqnarray}
This gives a condition $v_{\delta_R}/v_{\Delta_R}>1$.
{}From the bound on $m_{h}$ from $(\phi_2^0,\chi_1^0)$ submatrix,
we find
\begin{eqnarray}
0 < \frac 12 [-g_L^2(\kappa_2^2-\kappa_1^2)
-g_R^2(2v_{\Delta_R}^2-2v_{\delta_R}^2 -\kappa_2^2+\kappa_1^2)]
\end{eqnarray}
Thus the $D$-term, 
$D=2v_{\Delta_R}^2-2v_{\delta_R}^2+\kappa_2^2-\kappa_1^2$ has to be
negative. 
{}From the mass bound on the doubly charged Higgses, we get
\begin{eqnarray}
0&<&m_{H_1^{++}}^2\nonumber\\
&<&[-g_R^2(v_{\Delta_R}^2-v_{\delta_R}^2)
D]/
(v_{\Delta_R}^2+v_{\delta_R}^2)
+\frac 1M 8b_Rv_{\Delta_R}v_{\delta_R}\mu_{2R}
+\frac1{M^2}4b_R(2a_R+b_R)v_{\Delta_R}^2v_{\delta_R}^2.\nonumber\\
\label{mppbound}
\end{eqnarray}
We see that the $b_R$-parameter must necessarily be nonvanishing.

\begin{figure}[t]
\leavevmode
\begin{center}
\mbox{\epsfxsize=8.truecm\epsfysize=8.truecm\epsffile{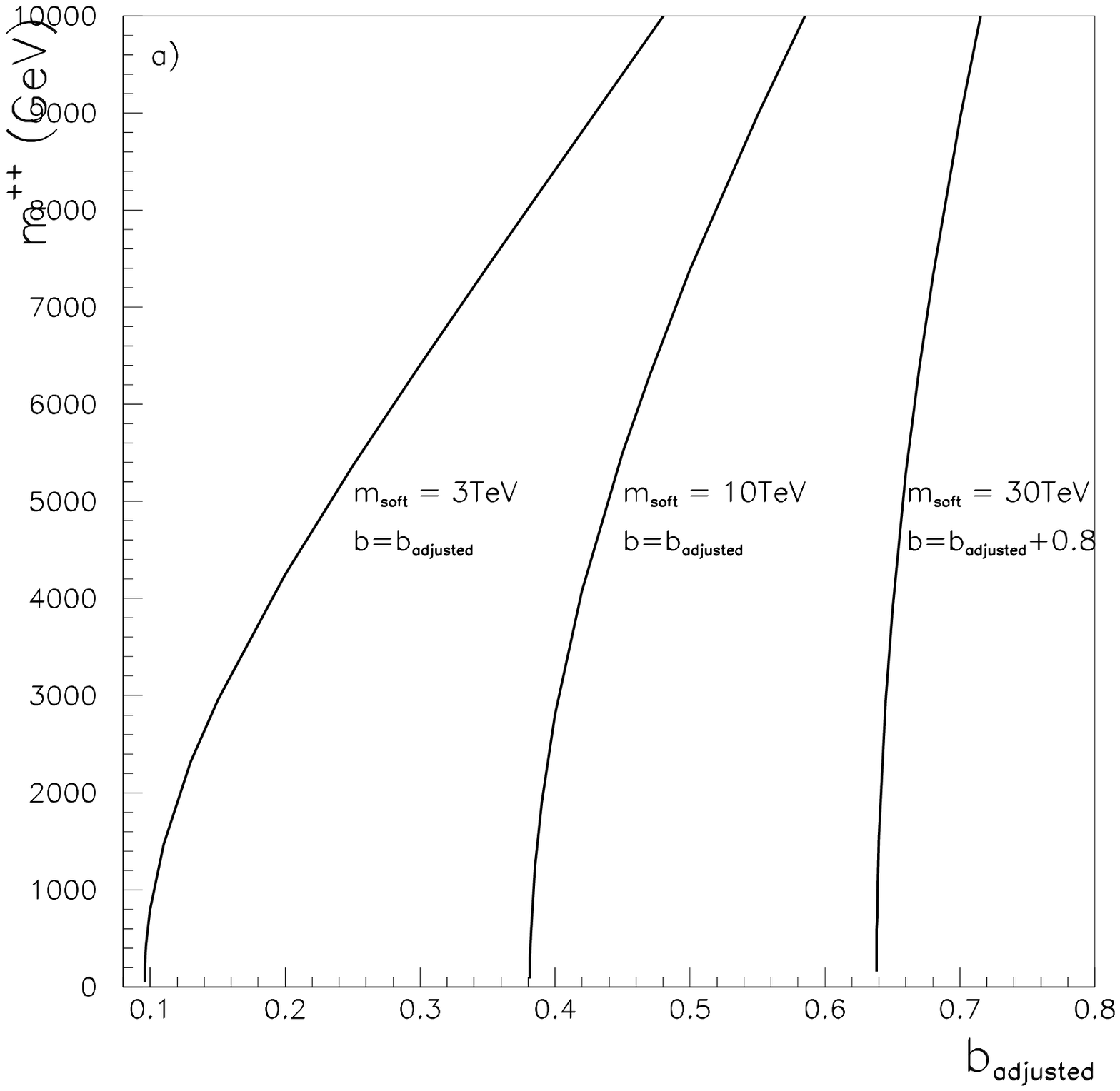}
\epsfxsize=8.truecm\epsfysize=8.truecm\epsffile{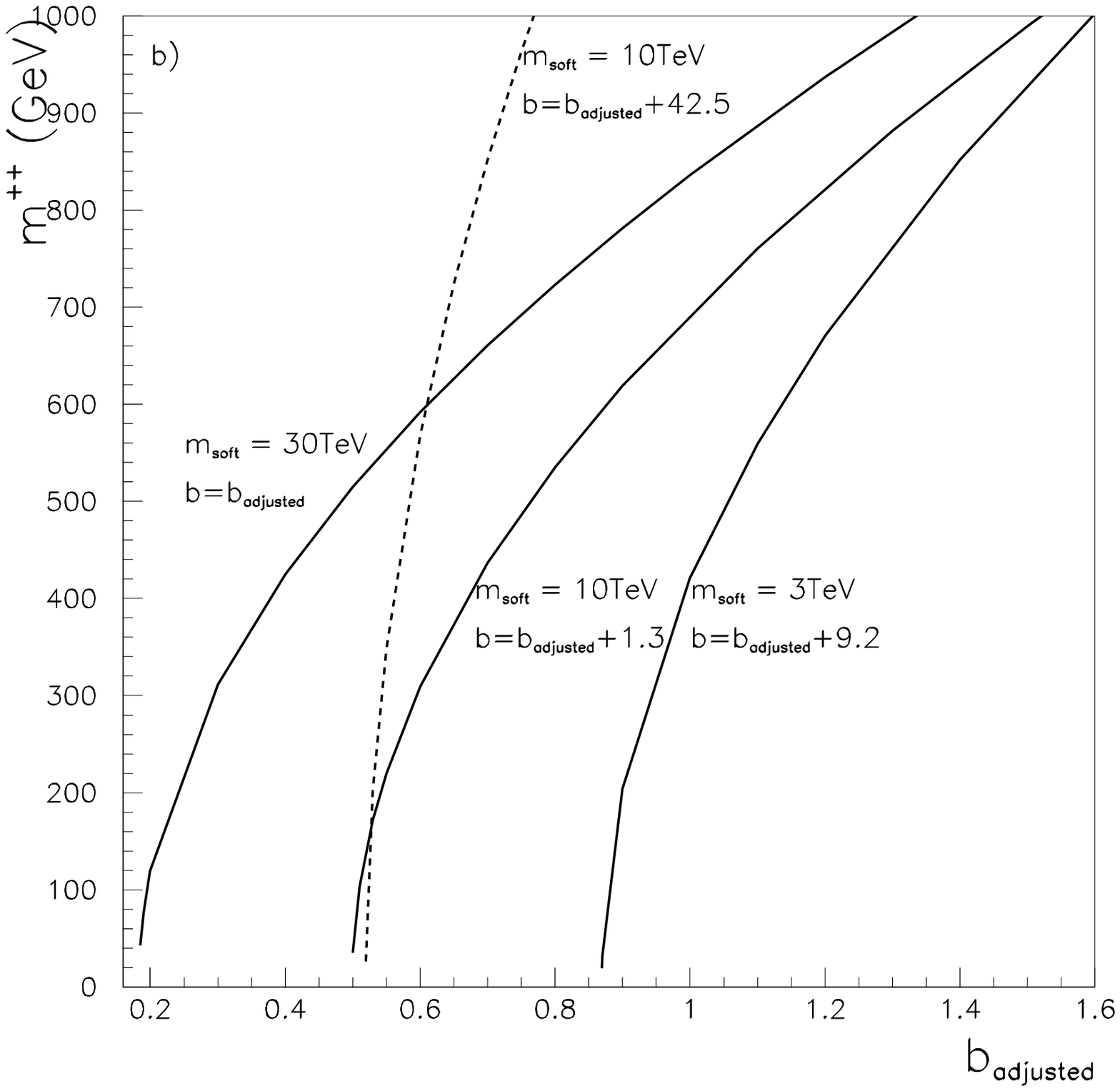}}
\end{center}
\caption{\label{nonren}
Mass of the doubly charged Higgs as a function of the
nonrenormalizable $b_R$-parameter.
In a) $v_R^2 / M = 10^4 $ GeV and 
$D  =m_{ soft }^2 $, while in b)
$v_R^2/M=10^{2}$ GeV and 
$D =(3$ TeV)$^2$ (solid line), except for 
$m_{soft} =10$ TeV also $D=(10$ TeV$^2$ is shown (dashed line).
The soft supersymmetry breaking parameters are marked in the figure,
as well as the $b_R$ parameters.
The curves are brought closer to each other for convenience.
In both figures $\tan\beta=50$, $M=10^{10}$ GeV, $\mu_{2R}=1$ TeV and
$\mu_1=\mu_1'=\mu_1''=500$ GeV.
In a) $M_R=10^7$ GeV and in b) $M_R=10^6$ GeV.}
\end{figure}

In Fig. \ref{nonren} we have plotted the mass of the lightest doubly 
charged Higgs boson
as a function of the nonrenormalizable $b_R$-parameter for two 
values of the ratios,
$v_R^2/M=10^2$ GeV and $v_R^2/M=10^4$ GeV.
One should note that at low energies $b_R$ is the only
nonrenormalizable parameter, which is necessarily nonvanishing.
In particular one can find solutions with $a_R=0$.
In Appendix B we give examples of mass spectra using the tree level
potential for the renormalizable and nonrenormalizable models.
One can expect that radiative corrections will change the spectrum
somewhat, but qualitative features should remain unchanged.
The value of $b_R$ needed to produce a proper minimum
depends on the other parameters in the model.
In Fig. \ref{nonren} a) $v_R^2/M=10^4$ GeV and $\sqrt{D}$ is the same as
the soft supersymmetry breaking parameters.
The mass of the flavour changing Higgs is roughly proportional to the value
of $\sqrt{D}$.
For larger values of the $D$-term, larger $b_R$'s are needed to retain positive
doubly charged Higgs mass, as seen from Eq. (\ref{mppbound}).
It is seen that the mass of the doubly charged Higgs rises rapidly  with
$b_R$ to multi-TeV region as is expected for large $v_R^2/M$.
In Fig.\ref{nonren} b) $v_R^2/M=10^2$ GeV,  
and $\sqrt{D}=3$ TeV (solid lines) for all 
the values of the soft supersymmetry breaking parameters that are plotted.
This shows the dependence on  the soft mass parameters.
For the soft mass parameter of 10 TeV we have plotted for comparison also the
dependence on $b_R$ when $\sqrt{D}=10$ TeV (dashed line).
It is seen that increasing the soft parameter, while the $D$-term 
remains constant, smaller $b_R$'s are needed to produce positive masses.
In Fig. (\ref{nonren}) we have taken  $\tan\beta$ to be 50.
Large $\tan\beta$ somewhat increases the masses of the neutral and
pseudoscalar Higgses and decreases the mass of the lightest doubly
charged Higgs.
With both values of $v_R^2/M$ it is possible to obtain a doubly charged
Higgs, which is light enough to be detected in future experiments,
but $m_{H^{++}}$ increases with $b_R$ much faster for large 
$v_R^2/M$.

The collider phenomenology of the doubly charged scalars has
been actively studied, since they appear in several extensions of the
Standard Model, can be relatively light and have clear signatures.
Here we will discuss the production of the doubly charged scalar
in colliders \cite{prod,HMPR}.
If the light doubly charged Higgs $H^{++}$ is lighter than any of the
supersymmetric partners,
the dominant decay mode of $H^{++}$ is to like sign leptons.

Kinematically,  production of a single doubly charged scalar would be
favoured.
If the couplings between electrons and triplet Higgses are 
large enough, the doubly charged particles can be
detected in $e^+e^-$ linear colliders in single production almost up
to the kinematical limit.
The $l^-l^-$ colliders are especially useful in studying the doubly
charged Higgs, since for nonzero triplet couplings it can be produced
as an s-channel resonance.
The possibilities for single production in hadron colliders in $WW$
fusion does not depend on the coupling to leptons, but it does depend,
in addition to $m_{H^{--}}$,  on the VEV of the triplet to which 
the doubly charged scalar belongs.
If $m_{W_R}\sim 1$ TeV, one can detect doubly charged Higgs bosons 
with $m_{H^{++}}\sim 1$ TeV at the LHC \cite{HMPR}.

The advantage of pair production of the doubly charged scalars 
compared to the single production is that
it is relatively model independent.
It can occur even if
$W_R$ is very heavy, as in the nonrenormalizable case, or the
triplet Yukawa couplings are very small. 
The doubly charged Higgses
can be produced in $f\bar f\rightarrow\gamma^*,Z^*\rightarrow
H^{++}H^{--}$ both at lepton and hadron colliders, if kinematically 
allowed.
At LHC this cross section falls off rapidly close to $m_{H^{++}}\sim
500$ GeV \cite{HMPR}.
The pair production cross section at Tevatron and at LEP II are given in
\cite{DM}.
For $e^+e^-$ linear colliders the pair production cross section for 
$\sqrt{s}=500$ GeV is $106\, (78)$ fb for $m_{H^{++}}\lsim 230\, (240)$ GeV.
Thus the detection of doubly charged scalars in pair production is
possible close to the kinematical limit.

\section{Conclusions}
\label{sec:conclusion}

Supersymmetric left-right models are well motivated extensions
of the MSSM, since they conserve $R$-parity as 
a consequence of gauge invariance.
We have made a detailed study of the Higgs sector in these models.
The lightest CP even Higgs boson can be considerably heavier than in the
MSSM, but its couplings to fermions remain similar to the couplings 
of the Standard Model Higgs boson.
If a Higgs, which nevertheless is too heavy to be the 
Higgs boson of the MSSM,   is detected, one should consider 
extended supersymmetric
models such as the ones studied in this paper.

In the SLRM with the minimal particle content one has typically also a
light doubly charged Higgs boson.
If this particle is found, it is a strong indication of the SLRM with
minimal particle content.

We wish to emphasize the importance of studying the full mass
matrices, including the soft mass parameters 
in determining the low energy mass spectrum, especially
the masses of the light scalar particles.

\section{Acknowledgments}

One of us (PNP) would like to thank the Helsinki Institute of Physics,
where this work was started, and to DESY, where it was completed,
for hospitality. The work of PNP is
supported by the University Grants Commission, India under
the project 10-26/98(SR-I).
The work of KH and KP is partially supported by the Academy of Finland
(no. 44129).

\appendix

\section{Appendix: Scalar potential of the minimal SLRM}
\label{sec:1}

The different components of the scalar potential 
(\ref{eq:scalarpotential}) for the minimal left-right 
supersymmetric model can be written as follows
($g_L$, $g_R$ and $g_{B-L}$ are the three gauge
couplings):
\begin{eqnarray}
V_F&=&
|h_{\Phi L} i\tau_2 \Phi L^c +h_{\chi L} i\tau_2 \chi L^c
+2h_{\delta_L}  L^{T}i\tau_2 \delta_L|^2 \nonumber\\
&&
+|h_{\Phi L} L^T i\tau_2\Phi + h_{\chi L} L^T i\tau_2\chi
+2h_{\Delta_R}  L^{cT}i\tau_2 \Delta_R |^2 \nonumber \\
&&
+|h_{\Delta_R} L^c  L^{cT} (i\tau_2) +\mu_{2R} \delta_R |^2
+|h_{\delta_L} L  L^{T} (i\tau_2) +\mu_{2L} \Delta_L |^2
\nonumber\\
&&
+|h_{\Phi Q}Q^cQ^T(i\tau_2)+h_{\Phi L}L^cL^T(i\tau_2)+
\mu_1(i\tau_2)\chi^T(i\tau_2) +2\mu'_1(i\tau_2)\Phi^T(i\tau_2) |^2
\nonumber\\
&&+
|h_{\chi Q}Q^cQ^T(i\tau_2)+h_{\chi L}L^cL^T(i\tau_2)+
\mu_1(i\tau_2)\Phi^T(i\tau_2) + 2\mu''_1(i\tau_2)\chi^T(i\tau_2)|^2
\nonumber\\
&& + |(i\tau_2)(h_{\Phi Q}\Phi +h_{\chi Q}\chi )Q^c|^2+
|Q^T(i\tau_2)(h_{\Phi Q}\Phi +h_{\chi Q}\chi )|^2\nonumber\\
&&+|\mu_{2R}\Delta_R|^2 +|\mu_{2L}\delta_L|^2,
\label{eq:fterms}
\end{eqnarray}
\begin{eqnarray}
V_D&=& \frac 18 g_L^2\sum_a \left[ \rm{Tr} (\Phi^\dagger\tau_a\Phi )+
\rm{Tr} (\chi^\dagger\tau_a\chi )
+2\rm{Tr} (\Delta_L^\dagger\tau_a\Delta_L )
+2\rm{Tr} (\delta_L^\dagger\tau_a\delta_L )\right.
\nonumber\\
&&\left. +L^\dagger\tau_a L+Q^\dagger\tau_a Q \right]^2
+\frac 18 g_R^2\sum_a \left[ -\rm{Tr} (\Phi\tau_a\Phi^\dagger )-
\rm{Tr} (\chi\tau_a\chi^\dagger )\right.  \nonumber\\
&&\left. +2\rm{Tr} (\Delta_R^\dagger\tau_a\Delta_R )
+2\rm{Tr} (\delta_R^\dagger\tau_a\delta_R )
+L^{c\dagger}\tau_a L^c+Q^{c\dagger}\tau_a Q^c \right]^2
\nonumber\\
&&+\frac 18 g_{B-L}^2\left[ 2\rm{Tr} (-\Delta_R^\dagger\Delta_R+
\delta_R^\dagger\delta_R-\Delta_L^\dagger\Delta_L+
\delta_L^\dagger\delta_L)\right.
\nonumber\\ &&
\left. -L^\dagger L+L^{c\dagger} L^c
+\frac 13 Q^\dagger Q-\frac 13 Q^{c\dagger} Q^c \right]^2,
\label{eq:dterms}
\end{eqnarray}
\begin{eqnarray}
V_{soft}&=&
m_\Phi^2\rm{Tr} |\Phi|^2 +  m_\chi^2\rm{Tr} |\chi|^2 -
(m_{\Phi \chi}^2  \rm{Tr} (i\tau_2 \Phi^T i\tau_2 \chi )
+m_{\Phi \Phi}^2  \rm{Tr} (i\tau_2 \Phi^T i\tau_2 \Phi )\nonumber\\
&&+m_{\chi \chi}^2  \rm{Tr} (i\tau_2 \chi^T i\tau_2 \chi ) +h.c.)
+m_{\Delta_R}^2 |\Delta_R |^2 + m_{\delta_R}^2|\delta_R |^2
-(m_{\Delta\delta}^2\rm{Tr} \Delta_R\delta_R +h.c.)\nonumber\\
&&+m_{\Delta_L}^2 |\Delta_L |^2 + m_{\delta_L}^2|\delta_L |^2
-({m_{\Delta\delta}}^{'2}\rm{Tr} \Delta_L\delta_L +h.c.)
+m_{L^c}^2 | L^c|^2
+m_L^2 |L|^2 \nonumber\\
&&+ (L^{T}i\tau_2 (A_\Phi\Phi +A_\chi\chi ) L^c
+A_{\Delta_R} L^{cT} i\tau_2 \Delta_R L^c
+A_{\delta_L} L^{T} i\tau_2 \delta_L L +h.c. )
\nonumber\\
&&+m_Q^2|Q|^2+ m_{Q^c}^2|Q^c|^2+
(Q^Ti\tau_2( h_{\Phi Q} A_{\Phi Q} \Phi +h_{\chi Q} A_{\chi Q} \chi )Q^c +h.c.).
\label{eq:softterms}
\end{eqnarray}
From
(\ref{eq:scalarpotential}), (\ref{eq:fterms}), (\ref{eq:dterms}) and
(\ref{eq:softterms}) it is straightforward to derive the mass matrix
for the CP-even Higgs scalars, whose eigenvalues will provide the
masses of the physical scalar Higgs bosons.

\section{Appendix: Examples of mass spectra}
\label{sec:2}

\begin{table}
\begin{tabular}{lll}
\hline
particle&mass (TeV) & composition\\
\hline
$H_{10}^0$&22.7 & $-0.3 \tilde\nu_R^r+0.7\Delta_R^{0r}-0.6\delta_R^{0r}$\\
$H_9^0$&20.2&$0.98\tilde\nu_L^r-0.2\phi_2^{0r}$\\
$H_8^0$&12.7&$\phi_1^{0r}$\\
$H_7^0$&11.9&$-0.2\tilde\nu^r-0.98\phi_2^{0r}+0.1\chi_1^{0r}$\\
$H_6^0$&10.3&$0.1\Delta_L^0-\delta_L^0$\\
$H_5^0$&9.70&$-\Delta_L^0-0.1\delta_L^0$\\
$H_4^0$&6.60&$-0.6\tilde\nu_R+0.3\Delta_R^0+0.7\delta_R^0$\\
$H_3^0$&3.53&$-0.1\phi_2^{0r}-\chi_1^{0r}$\\
$H_2^0$&1.95&$0.7\tilde\nu_R+0.6\Delta_R^0+0.4\delta_R^0$\\
$H_1^0$&0.096&$\chi_2^{0r}$\\
$A_8$&22.3&$-0.9 \tilde\nu_R^i-0.5\Delta_R^{0i}-0.2\delta_R^{0i}$\\
$A_7$&20.2&$0.98\tilde\nu_L^i+0.2\phi_2^{0i}$\\
$A_6$&12.7&$\phi_1^{0i}$\\
$A_5$&12.4&$0.5\tilde\nu_R^i-0.5\Delta_R^{0i}-0.7\delta_R^{0i}$\\
$A_4$&11.9&$0.2\tilde\nu_L^i-0.98\phi_2^{0i}-0.1\chi_1^{0i}$\\
$A_3$&10.3&$-0.1\Delta_L^{0i}-\delta_L^{0i}$\\
$A_2$&9.70&$-\Delta_L^{0i}-0.1\delta_L^{0i}$\\
$A_1$&3.53&$0.1\phi_2^{0i}-\chi_1^{0i}$\\
$H_8^+$&20.2&$0.98\tilde e_L+0.2\phi_1^+$\\
$H_7^+$&19.0&$-0.4\tilde e_R+0.7\Delta_R^+-0.5\delta_R^+$\\
$H_6^+$&12.7&$\phi_2^+$\\
$H_5^+$&11.9&$-0.2\tilde e_L+0.98\phi_1^+$\\
$H_4^+$&10.3&$0.1\Delta_L^{+}-\delta_L^{+}$\\
$H_3^+$&9.70&$-\Delta_L^{+}-0.1\delta_L^{+}$\\
$H_2^+$&9.42&$-0.8\tilde e_R-0.1\Delta_R^++0.5\delta_R^+$\\
$H_1^+$&3.53&$-0.1\phi^+_1+\chi_2^+$\\
$H_4^{++}$&14.7&$0.8\Delta_R^{++}-0.6\delta_R^{++}$\\
$H_3^{++}$&10.3&$0.1\Delta_L^{++}-\delta_L^{++}$\\
$H_2^{++}$&9.70&$-\Delta_L^{++}-0.1\delta_L^{++}$\\
$H_1^{++}$&0.169&$-0.6\Delta_R^{++}-0.8\delta_R^{++}$\\
\hline
\end{tabular}
\caption{Mass spectrum of a renormalizable model with minimum particle
content.\label{table1}}
\end{table}

\begin{table}
\begin{tabular}{lll}
\hline
particle&mass (TeV) & composition\\
\hline
$H_8^0$&1570&$0.7(\delta_R^{0r}-\Delta_R^{0r})$\\
$H_7^0$&53.8&$0.1\phi_1^{0r}+0.3\phi_2^{0r}-0.9\chi_2^{0r}$\\
$H_6^0$&52.9&$0.9\phi_2^{0r}-0.1\chi_1^{0r}+0.3\chi_2^{0r}$\\
$H_5^0$&11.7&$\Delta_L^{0r}$\\
$H_4^0$&9.24&$-0.1\phi_2^{0r}-\chi_1^{0r}$\\
$H_3^0$&8.0&$\delta_L^{0r}$\\
$H_2^0$&0.284&$-0.7(\delta_R^{0r}+\Delta_R^{0r})$\\
$H_1^0$&0.091&$\phi_1^{0r}+0.1\chi_2^{0r}$\\
$A_6^0$&53.8&$0.1\phi_1^{0i}+0.3\phi_2^{0i}+0.9\chi_2^{0i}$\\
$A_5^0$&52.9&$-0.9\phi_2^{0i}-0.1\chi_1^{0i}+0.3\chi_2^{0i}$\\
$A_4^0$&14.1&$-0.7(\delta_R^{0i}+\Delta_R^{0i})$\\
$A_3^0$&11.7&$\Delta_L^{0i}$\\
$A_2^0$&9.24&$0.1\phi_2^{0i}-\chi_1^{0i}$\\
$A_1^0$&8.0&$\delta_L^{0i}$\\
$H_6^+$&929&$0.7(\delta_R^{+}-\Delta_R^{+})$\\
$H_5^+$&53.8&$-0.1\chi_1^+-0.3\chi_2^+-0.9\phi_2^+$\\
$H_4^+$&52.9&$-0.9\chi_2^+-0.1\phi_1^++0.3\phi_2^+$\\
$H_3^+$&11.7&$\Delta_L^+$\\
$H_2^+$&9.24&$-0.1\chi_2^++\phi_1^+$\\
$H_1^+$&8.0&$\delta_L^+$\\
$H_4^{++}$&19.6&$-0.98\delta_R^{++}+0.2\Delta_R^{++}$\\
$H_3^{++}$&11.7&$\Delta_L^{++}$\\
$H_2^{++}$&8.0&$\delta_L^{++}$\\
$H_1^{++}$&0.202&$-0.2\delta_R^{++}-0.98\Delta_R^{++}$\\
\hline
\end{tabular}
\caption{Mass spectrum of a nonrenormalizable model with minimum particle
content.\label{table2}}
\end{table}

Here we give one example each of the 
mass spectrum of Higgs bosons in  renormalizable
and nonrenormalizable suoersymmetric left-right models.
In table \ref{table1} we have shown the spectrum 
in the renormalizable model, and 
in table \ref{table2} the spectrum for the nonrenormalizable 
model with minimal particle content.
In both examples the soft supersymmetry breaking parameters are
$\sim 10$ TeV.
In table \ref{table1}, the right handed scale is 10 TeV, and
in table \ref{table2} $v_R^2/M=100$ GeV.
We have  denoted  $h=H_1^0$ and $H=H_2^0$.

\newpage




\newpage

\end{document}